\documentclass[12pt]{iopart}

\pdfoutput=1

\usepackage[english]{babel}
\usepackage{
amssymb,amsfonts,latexsym}
\usepackage{graphicx}
\usepackage{color}
\usepackage{iopams}
\usepackage{subfigure}
\usepackage{afterpage}
\usepackage{blkarray}
\usepackage{cite}
\usepackage{tikz}
\usetikzlibrary{snakes}
\usetikzlibrary{calc}
\usepackage{bm}
\usepackage{braket}

\definecolor{blue(pigment)}{rgb}{0.2, 0.2, 0.6}
\usepackage{mathrsfs}
\usepackage{times}
\usepackage[colorlinks,
citecolor=blue,linkcolor=blue,urlcolor=blue]{hyperref}
\usepackage{scalerel}
\usepackage{tensor}
\usepackage{etoolbox}

\makeatletter
\def\@mkboth#1#2{}
\newlength\appendixwidth
\preto\appendix{\addtocontents{toc}{\protect\patchl@section}}
\newcommand{\patchl@section}{%
  \settowidth{\appendixwidth}{\textbf{Appendix }}%
  \addtolength{\appendixwidth}{1.5em}%
  \patchcmd{\l@section}{1.5em}{\appendixwidth}{}{\ddt}%
}
\makeatother

\def\eqref#1{(\ref{#1})}

\usepackage{cancel}

\newcommand{\be}{\begin{equation}}
\newcommand{\ee}{\end{equation}}
\newcommand{\bea}{\begin{eqnarray}}
\newcommand{\eea}{\end{eqnarray}}

\newcommand\reallywidehat[1]{\arraycolsep=0pt\relax%
\begin{array}{c}
\stretchto{
  \scaleto{
    \scalerel*[\widthof{\ensuremath{#1}}]{\kern-.5pt\bigwedge\kern-.5pt}
    {\rule[-\textheight/2]{1ex}{\textheight}} 
  }{\textheight} %
}{0.5ex}\\           
#1\\                 
\rule{-1ex}{0ex}
\end{array}
}

\begin{document}

\title[Integrable quenches in nested spin chains II]{Integrable quenches in nested spin chains II:\\ 
fusion of boundary transfer matrices}
\author{Lorenzo Piroli$^{1,2}$, Eric Vernier$^{3}$, Pasquale Calabrese$^{1,4}$, Bal\'{a}zs Pozsgay$^{5,6}$}
\address{$^1$ SISSA and INFN, via Bonomea 265, 34136 Trieste, Italy}
\address{$^2$ Max-Planck-Institut f\"ur Quantenoptik, Hans-Kopfermann-Str. 1, 85748 Garching, Germany}
\address{$^3$ The Rudolf Peierls Centre for Theoretical Physics, Oxford University, Oxford, OX1 3NP, United Kingdom.}
\address{$^4$ International Centre for Theoretical Physics (ICTP), I-34151, Trieste, Italy}
\address{$^5$ Department of Theoretical Physics, Budapest University
	of Technology and Economics, 1111 Budapest, Budafoki \'{u}t 8, Hungary}
\address{$^6$ BME Statistical Field Theory Research Group, Institute of Physics,
	Budapest University of Technology and Economics, H-1111 Budapest, Hungary}
\date{\today}

\begin{abstract}
	We consider quantum quenches in the integrable $SU(3)$-invariant spin chain (Lai-Sutherland model), and focus on the family of integrable initial states. By means of a Quantum Transfer Matrix approach, these can be related to ``soliton-non-preserving'' boundary transfer matrices in an appropriate transverse direction. In this work, we provide a technical analysis of such integrable transfer matrices. In particular, we address the computation of their spectrum: this is achieved by deriving a set of functional relations between the eigenvalues of certain ``fused operators'' that are constructed starting from the soliton-non-preserving boundary transfer matrices (namely the $T$- and $Y$-systems).  As a direct physical application of our analysis, we compute the Loschmidt echo for imaginary and real times after a quench from the integrable states. Our results are also relevant for the study of the spectrum of $SU(3)$-invariant Hamiltonians with open boundary conditions.
\end{abstract}


\maketitle

{\hypersetup{linkcolor=blue(pigment)}
\tableofcontents
}

\section{Introduction}
\label{sec:intro}

This is the second of two works devoted to the study of quantum
quenches in the $SU(3)$-invariant Lai-Sutherland chain
\cite{lai-74,sutherland-75}
\be
H_L=\sum_{j=1}^{L}\left[{\bf s}_j\cdot {\bf s}_{j+1}+\left({\bf s}_j\cdot {\bf s}_{j+1}\right)^2\right]-2L\,,
\label{eq:hamiltonian}
\ee
where the spin-$1$ operators $s^{a}_j$ are given by the standard three-dimensional representation of the $SU(2)$ generators. This integrable model is interesting from the physical point of view, as it displays different quasi-particle species, each one forming an infinite number of bound states. While many properties of the system at zero and finite temperature are by now well understood \cite{jls-89,mntt-93,johannesson-86,johannesson2-86,dn-98,Tsub03}, including the knowledge of its correlation functions \cite{FuKl99,RiKl18}, until recently the analytical study of quench problems in this model has been out of our reach. The main reason for this lies in the fact that the solution to the Hamiltonian \eqref{eq:hamiltonian} involves a complicated \emph{nested} Bethe ansatz \cite{kr-81,efgk-05}, for which generalizations of recent analytic advances in integrability out of equilibrium \cite{CaEM16} are not straightforward, including the string-charge duality \cite{IDWC15,IQDB16,IMPZ16,PiVC16,IlQC17} or the Quench Action method \cite{CaEs13,Caux16,DWBC14,WDBF14,PMWK14}. Indeed, until very recently no initial state was known for which an explicit characterization of the corresponding post-quench steady state could be achieved. We note that, while it is established that for non-nested integrable systems the latter always corresponds to a generalized Gibbs ensemble (GGE) \cite{RDYO07,ViRi16,EsFa16,FaEs13,Pozs13_GGE,PVCR17,FCEC14,PoVW17}, the picture is less clear in nested systems, where even a satisfactory understanding of the structure of quasi-local conservation laws is missing.

In the first of our works \cite{PVCP18}, it was shown that one can successfully tackle the study of quantum quenches in the nested chain \eqref{eq:hamiltonian} by means of an approach based on the Quantum Transfer Matrix (QTM)  formalism \cite{Klum92,Klum04}. The latter is complementary to the string-charge duality and the Quench Action method, and was initially introduced for non-nested $XXZ$ Heisenberg chains in \cite{PiPV17} (see also \cite{Pozs13}). The starting point of this approach is to identify a family of initial integrable states \cite{PiPV17_2} which, in close analogy with classical results in the context of quantum field theory \cite{GhZa94,Delf14,Schu15}, can be related to integrable boundary conditions in an appropriate transfer direction (or channel). More precisely, to each initial integrable state one can associate a given transfer matrix with open boundary conditions; in the case of the $SU(3)$-invariant Hamiltonian \eqref{eq:hamiltonian} these were shown to correspond to the ``soliton-non-preserving'' case studied in \cite{Doik00,AACD04,AACD04_2}. 

Importantly, the steady state reached at large times after a quench from integrable states could be characterized in \cite{PVCP18} in terms of the corresponding quasi-particle distribution functions, which were computed analytically. This result is non-trivial: indeed, except for the special matrix product states found in \cite{LeKM16,LeKL18} , whose quench dynamics was considered in \cite{MBPC17}, no other example is known where the post-quench steady state can be determined, even approximately, for nested Hamiltonians.
The derivation of \cite{PVCP18} was based on a comparison between the QTM and Quench Action methods, and exploited as a technical ingredient certain ``fusion relations'' of the $SU(3)$-invariant boundary transfer matrices associated with the integrable states. These relations generalize to the nested case the ones derived in \cite{Zhou95, Zhou96} for the $SU(2)$-invariant case (and the corresponding $q$-deformations).

Albeit important, the derivation of such fusion relations is very technical, and was beyond the scope of Ref.~\cite{PVCP18}. This task is thus carried out in this more specialized work, where the fusion relations (and the corresponding $Y$-system \cite{KuNS11}) are derived and analyzed in detail. In turn, these findings in principle give us access to the full spectrum of the boundary transfer matrix \cite{PiPV18}, and are thus relevant also for the study of $SU(3)$-invariant Hamiltonians with open boundary conditions. As an immediate physical application in the context of non-equilibrium physics, we address the computation of the Loschmidt echo \cite{qslz-06} after a quench from integrable states, which is both interesting at real times and imaginary times, due to its connection to the study of dynamical quantum phase transitions \cite{hpk-13,pmgm-10,fagotti2-13,dpfz-13,ks-13,ks-17,
kks-17,cwe-14,heyl-14,vths-13,as-14,deluca-14,heyl-15,
kk-14,vd-14,ssd-15,JK-15,AbK-16,sdpd-16,zhks-16,zy-16,ps-16,Heyl17,
JaJo17,HaZa17,ZaHa17} and the statistics of the work performed by the quench \cite{silva-08,ps-14}, respectively. We work it out exactly at imaginary time and for small real times.  Altogether, the calculations presented in \cite{PVCP18} and in this work, provide a comprehensive analysis of integrable quenches in the $SU(3)$-invariant spin chain \eqref{eq:hamiltonian}. Our findings are likely to have ramifications in the study of other nested integrable systems such as, for instance, multi-component Fermi and Bose ultra-cold gases, which are of special relevance for cold-atom experimental realizations \cite{BlDZ08,PMCL14,GuBL13}. 

The organization of this article is as follows. In Sec.~\ref{sec:general_setting} we introduce the algebraic Bethe ansatz, and recall the construction of the integrable states carried out in \cite{PVCP18}. We introduce in particular the corresponding soliton-non-preserving boundary transfer matrices, whose fusion relations are derived explicitly in Sec.~\ref{sec:fusion_relations}. In Sec.~\ref{sec:Loschmidt} we use them to compute the Loschmidt echo, both at imaginary and real times. Our conclusions are consigned to Sec.~\ref{sec:conclusions}.

\section{The general setting}\label{sec:general_setting}

\subsection{The periodic algebraic Bethe ansatz}
We begin by recalling the technical tools which are needed in order to analyze the Hamiltonian~\eqref{eq:hamiltonian}. In \cite{PVCP18}, we reviewed the main aspects of the coordinate
nested Bethe ansatz, with which one is able to directly write down the
wave functions corresponding to the eigenstates of the Hamiltonian
\eqref{eq:hamiltonian}. While this approach provides a clear physical picture, a quantitative analysis is better carried out by means of its algebraic version, which we briefly review. The main object of the theory is the $R$-matrix $\widehat{R}_{12}(\lambda)$ which acts on the tensor product $h_1\otimes h_2$, with $h_j\simeq \mathbb{C}^3$. Explicitly, it reads
\begin{equation}
\widehat{R}_{12}(\lambda)=\frac{1}{\lambda+i}\left(\lambda + i \mathcal{P}_{12}\right)\,,
\end{equation}
where $\mathcal{P}_{12}$ is the permutation matrix exchanging the spaces $h_1$ and $h_2$
\be
\mathcal{P}_{12}|a\rangle_1\otimes |b\rangle_{2}=|b\rangle_1\otimes |a\rangle_{2}\,,
\label{eq:permutation_matrix}
\ee
and which can be written in terms of the spin-1 operators as $\mathcal{P}_{12} = -1 + \vec{S}_1 \cdot \vec{S}_2 + (\vec{S}_1 \cdot \vec{S}_2)^2$.
The $R$-matrix satisfies the following properties of regularity and unitarity
\bea
\widehat{R}_{12}(0)=\mathcal{P}_{12}\,,\\
\widehat{R}_{12}(u)\widehat{R}_{21}(-u)=1\,,
\eea
where
\be
\widehat{R}_{21}(u)=\mathcal{P}_{12}\widehat{R}_{12}(u)\mathcal{P}_{12}\,.
\ee
Note that one has simply $\widehat{R}_{12}(u)=\widehat{R}_{21}(u)$. We also introduce the $R$-matrix with a different normalization
\be
R_{12}(\lambda)=(\lambda+i)\widehat{R}_{12}(\lambda)=\lambda+i\mathcal{P}_{12}\,,
\ee
which satisfies
\bea
\label{inversion1}
R_{12}(\lambda)R_{21}(-\lambda)=\zeta(\lambda)\,,\\
\label{inversion2}
R^{t_1}_{12}(\lambda)R^{t_1}_{21}(-\lambda-3i)=\bar{\zeta}(\lambda+3i/2)\,,
\label{inversion2t}
\eea
where 
\bea
\zeta(\lambda)=(\lambda+i)(-\lambda+i)\,,\qquad \bar{\zeta}(\lambda)=(\lambda+3i/2)(-\lambda+3i/2)\,.
\eea
Here we introduced the transposition
\be
A^{t}=V^{-1}A^{T}V\,,
\ee
where $A^{T}$ is the usual transposed of the matrix $A$, while
\be
V=\left(
\begin{array}{ccc}
	0 & 0 & 1 \\
	0 & 1 & 0 \\
	1 & 0 & 0
\end{array}
\right)\,.
\ee
Moreover, the notation $A^{t_1}$ in  \eqref{inversion2t} signals the partial transposition, which is performed only on the indices refering to $h_1$ in the tensor product $h_1 \otimes h_2$.
In the following, it will be also necessary to introduce the $R$-matrix involving the conjugate representation of $SU(3)$ \cite{VeWo92,AbRi96,Doik00_f}. The latter is given as follows
\be
\label{Rbar}
\bar{R}_{12}(\lambda):=R_{12}^{t_1}\left(-\lambda-\frac{3i}{2}\right)\,.
\ee
The $R$-matrix $\bar{R}_{12}(\lambda)$ can be interpreted physically as the scattering matrix describing the interaction between a soliton and an anti-soliton, and fulfills
\bea
\bar{R}_{12}(\lambda)\bar{R}_{21}(-\lambda)=\bar{\zeta}(\lambda)\,,\\
\bar{R}^{t_1}_{12}(\lambda)\bar{R}^{t_1}_{12}(-\lambda-3i)=\zeta(\lambda+3i/2)\,.
\eea

The $R$-matrices introduced above satisfy the Yang-Baxter equations, which take the form
\bea
R_{12}(\lambda)R_{13}(\lambda+\mu)R_{23}(\mu)=R_{23}(\mu)R_{13}(\lambda+\mu)R_{12}(\lambda) \,,\label{eq:YBER11}\\
\bar{R}_{12}(\lambda)\bar{R}_{13}(\lambda+\mu)R_{23}(\mu)=R_{23}(\mu)\bar{R}_{13}(\lambda+\mu)\bar{R}_{12}(\lambda)\,.
\label{eq:YBER12}
\eea
These fundamental relations have several immediate consequences. Indeed, defining the transfer matrices acting on the global Hilbert space $\mathcal{H}=h_1\otimes \ldots \otimes h_N$ as
\bea
t(\lambda)&=&{\rm tr}_j\left[\widehat{R}_{jL}(\lambda-\xi_L)\ldots \widehat{R}_{j1}(\lambda-\xi_1)\right]\,,\label{eq:periodic_transfer_matrix}\\
\bar{t}(\lambda)&=&{\rm tr}_j\left[\widehat{R}_{1j}(\lambda+\xi_1)\ldots \widehat{R}_{Lj}(\lambda+\xi_L)\right]\label{eq:periodic_bar_transfer_matrix}\,,
\eea
it follows that 
\be
\left[t(\lambda),t(\mu)\right]=\left[\bar{t}(\lambda),t(\mu)\right]=\left[\bar{t}(\lambda),\bar{t}(\mu)\right]=0\,.
\label{eq:commutation_relations_t}
\ee
The traces in \eqref{eq:periodic_transfer_matrix}, \eqref{eq:periodic_bar_transfer_matrix} are taken over the auxiliary space $h_j\simeq \mathbb{C}^3$. The importance of this relation becomes manifest when complemented with the following trace formula 
\be
H_L=i\frac{\partial}{\partial\lambda}\ln t(\lambda)\Bigr|_{\lambda=0}\,.
\label{eq:charges_hamiltonian_definition}
\ee 
From \eqref{eq:commutation_relations_t} and \eqref{eq:charges_hamiltonian_definition} it is clear that in order to diagonalize the Hamiltonian $H_L$ it is sufficient to find the eigenspectrum of $t(\lambda)$. The algebraic Bethe ansatz directly provides us with the tools to do exactly this. By a standard procedure, it is possible to prove \cite{kr-81} that the eigenvalues of $t(\lambda)$ are labeled by two sets of rapidities $\{k_j\}_{j=1}^N$, $\{\lambda_j\}_{j=1}^M$ satisfying the Bethe equations
\bea
\left(\frac{k_{j}+i/2}{k_{j}-i/2}\right)^{L}=\prod_{\scriptstyle p=1\atop \scriptstyle p\ne j}^{N} \frac{k_j- k_p+ i}{k_j- k_p- i} \prod_{\ell = 1}^{M} \frac{\lambda_\ell- k_j+ i/2}{\lambda_\ell- k_j- i/2}\,, \quad j=1,\ldots,N\,,\label{eq:bethe_equations1}\\
1=\prod_{j=1}^N \frac{k_j-\lambda_\ell- i/2}{k_j-\lambda_\ell+ i/2}\prod_{\scriptstyle m=1\atop \scriptstyle m\neq \ell}^M  \frac{\lambda_\ell-\lambda_m- i}{\lambda_\ell-\lambda_m+ i}\,, \quad \ell=1,\ldots,M\,.
\label{eq:bethe_equations2}
\eea
In particular, by denoting with $\ket{\{k_j\},\{\lambda_j\}}$ the corresponding eigenvectors we have
\be
t(\lambda)\ket{\{k_j\},\{\lambda_j\}}=\nu(\{k_j\},\{\lambda_j\},\lambda)\ket{\{k_j\},\{\lambda_j\}}\,,
\label{eq:transfer_matrix_eigenvalue}
\ee
where 
\bea
 \nu\left(\{k_j\},\{\lambda_j\},\lambda\right)=\left[a(\lambda)\right]^L\prod_{j=1}^{N}\frac{1}{a(\lambda-k_j+i/2)}\nu_1\left(\{k_j\},\{\lambda_j\},\lambda\right)\nonumber\\
+\prod_{j=1}^N\frac{1}{a(k_j-i/2-\lambda)}\,,
\label{eq:transfer_eigenvalue}
\eea
with the additional definitions
\bea
\fl \nu_1(\{k_j\},\{\lambda_j\},\lambda)=\prod_{j=1}^Na(\lambda-k_j+i/2) \prod_{r=1}^{M}\frac{1}{a(\lambda-\lambda_r+i/2)}+\prod_{r=1}^{M}\frac{1}{a(\lambda_r-i/2-\lambda)}\,,
\eea
and
\bea
a(\lambda)=\frac{\lambda}{\lambda+i}\,.
\eea

\subsection{The boundary algebraic Bethe ansatz}

The transfer matrices $t(\lambda)$ and $\bar{t}(\lambda)$ defined in \eqref{eq:periodic_transfer_matrix} and \eqref{eq:periodic_bar_transfer_matrix} are translationally invariant operators corresponding to periodic boundary conditions. It is possible to define analogous operators in the case where open boundaries are assumed, via the boundary algebraic Bethe ansatz \cite{Skly88}. As we have already seen in \cite{PVCP18}, in the case of higher rank algebras $SU(\mathcal{N})$ with $\mathcal{N}\geq 3$, there exist two inequivalent boundary transfer matrices, which correspond to the so-called soliton-preserving \cite{dn-98} and soliton-non-preserving \cite{Doik00} boundary conditions \cite{AACD04}. In this work we will be interested in the latter, which are reviewed in the following.

We recall that the soliton-non-preserving boundary transfer matrices acting on a chain of length $N$, composed of an alternating product of fundamental and conjugate representations, are defined as
\bea
\tau(\lambda)&=&{\rm tr}_{a}\left\{K_a^{+}(\lambda)T_a(\lambda)K_a^{-}(\lambda)\hat{T}_{\bar{a}}(\lambda)\right\}\,,\label{eq:def_tau}\\
\bar{\tau}(\lambda)&=&{\rm tr}_a\left\{K_{\bar{a}}^{+}(\lambda)T_{\bar{a}}(\lambda)K_{\bar{a}}^{-}(\lambda)\hat{T}_a(\lambda)\right\}\,,\label{eq:def_bar_tau}
\eea
where the following definitions are used
\bea
T_a(\lambda)=R_{aN}(\lambda-\xi_{N})\bar{R}_{a(N-1)}(\lambda-\xi_{N-1})\ldots R_{a2}(\lambda-\xi_2) \bar{R}_{a1}(\lambda-\xi_1)\,,\label{eq:def_t_a}\\
\hat{T}_{\bar{a}}(\lambda)=R_{1a}(\lambda+\xi_1)\bar{R}_{2a}(\lambda+\xi_2)\ldots R_{(N-1)a}(\lambda+\xi_{N-1}) \bar{R}_{Na}(\lambda+\xi_{N})\,,\\
T_{\bar{a}}(\lambda)=\bar{R}_{aN}(\lambda-\xi_{N})R_{a(N-1)}(\lambda-\xi_{N-1})\ldots \bar{R}_{a2}(\lambda-\xi_2) R_{a1}(\lambda-\xi_1)\,,\label{eq:def_t_bar_a}\\
\hat{T}_{a}(\lambda)=\bar{R}_{1a}(\lambda+\xi_1)R_{2a}(\lambda+\xi_2)\ldots \bar{R}_{(N-1)a}(\lambda+\xi_{N-1}) R_{Na}(\lambda+\xi_{N})\,.
\eea
Here we introduced the free parameters $\xi_j$ (called inhomogeneities) while the trace in \eqref{eq:def_tau} and \eqref{eq:def_bar_tau} is taken over the auxiliary space $h_a\simeq \mathbb{C}^3$. Finally, $K_a^{\pm}(\lambda)$, $K_{\bar{a}}^{\pm}(\lambda)$ are $3\times 3$ matrices.

Importantly, the $K$-matrices $K_a^{\pm}(\lambda)$, $K_{\bar{a}}^{\pm}(\lambda)$ have to be chosen in such a way that transfer matrices with different spectral parameter commute. This can be done by setting $K_{a}^-(\lambda)=K_{a}(\lambda)$, $K_{\bar{a}}^-(\lambda)=K_{\bar{a}}(\lambda)$,  $K_{a}^+(\lambda)=K^t_{a}(-\lambda-i3/2)$ and $K_{\bar{a}}^+(\lambda)=K^t_{\bar{a}}(-\lambda-i3/2)$, where $K_a(\lambda)$ and $K_{\bar{a}}(\lambda)$  are a solution to the following twisted boundary Yang-Baxter (or reflection) equations 
\bea
\fl R_{ab}(\lambda-\mu)K_a(\lambda)\bar{R}_{ba}(\lambda+\mu)K_b(\mu)=K_b(\mu)\bar{R}_{ab}(\lambda+\mu)K_a(\lambda)R_{ba}(\lambda-\mu)\,,\label{eq:reflection_1}\\
\fl \bar{R}_{ab}(\lambda-\mu)K_{\bar{a}}(\lambda)R_{ba}(\lambda+\mu)K_b(\mu)=K_b(\mu)R_{ab}(\lambda+\mu)K_{\bar a}(\lambda){\bar R}_{ba}(\lambda-\mu)\,.\label{eq:reflection_2}
\eea
A classification of all invertible $K$-matrices satisfying the reflection equations was performed in \cite{AACD04}. In particular, it was shown that for the $SU(3)$-case the only invertible solutions to the latter equations are scalar matrices $K^-(\lambda)=K^-$ such that $(K^-)^{t} = K^-$, namely such that $\tilde{K}^- = K^- V$ is symmetric. The most general matrix of this form is written as
\be
K_a(\lambda)=K_sV\,,
\label{eq:K_symmetric}
\ee
where $K_{s}$ a symmetric numerical matrix
\be
K_{s}=\left(
\begin{array}{ccc}
	\kappa_{11} & \kappa_{12} & \kappa_{13}\\
	\kappa_{12} & \kappa_{22} & \kappa_{23}\\
	\kappa_{13} & \kappa_{23} & \kappa_{33}
\end{array}
\right)\,,
\label{eq:k_s}
\ee
with ${\rm det}K_s\neq 0$.

The classification performed in \cite{AACD04} does not treat
  non-invertible matrices. It follows from continuity
  that any symmetric matrix with ${\rm det}K_s=0$ also satisfies the
  reflection equations, but here we discard these solutions. On the one hand, the fusion relations to be treated
  below depend crucially on a non-vanishing determinant. On the other hand, solutions with ${\rm det}K_s=0$ can be rotated to lie in an $SU(2)$ sector, and thus the earlier results in the literature can  be used to solve these cases (see also Ref.~\cite{PVCP18}).

\subsection{The integrable states}
\label{sec:integrable_states}

As we have seen in \cite{PVCP18}, there is a close relation between the soliton-non-preserving boundary transfer matrices introduced above and a special class of product states, namely the integrable states. We recall that the latter were initially introduced in \cite{PiPV17_2} as those matrix product states \cite{PVWC06} (with a finite bond-dimension) which are annihilated by all the parity-odd conserved charges of the Hamiltonian. In the case of the $XXZ$ Heisenberg chain, it was shown how to derive an infinite family of such states from integrable boundary conditions in an appropriate rotated channel. The constructions of \cite{PiPV17_2} were generalized to the $SU(3)$-invariant spin chain \eqref{eq:hamiltonian} in \cite{PVCP18}, and a class of integrable product states was explicitly derived.

The family of product states found in \cite{PVCP18} reads
\be
|\Psi_0\rangle=|\psi_0\rangle_{1,2}\otimes\ldots \otimes |\psi_0\rangle_{L-1,L}\,.
\label{eq:initial_state}
\ee
where the two-site state has to be chosen as
\bea 
| \psi_0 \rangle &=& \kappa_{11} |1,1\rangle +\kappa_{22} |2,2\rangle +  \kappa_{33} |3,3\rangle 
+ \kappa_{12}( |1,2\rangle +  |2,1\rangle ) \nonumber\\
&+& \kappa_{13}( |1,3\rangle +  |3,1\rangle ) 
+ \kappa_{23}( |2,3\rangle +  |3,2\rangle )\,.
\label{eq:integrable_block}
\eea  
Here we denoted with $|1\rangle$, $|2\rangle$ and $|3\rangle$ the basis vectors of the locals spaces $h_j\simeq \mathbb{C}^3$ and used the convention
\be
|\alpha_1\,,\alpha_2\,,\ldots \alpha_L\rangle=|\alpha_1\,\rangle\otimes |\alpha_2\rangle\otimes \ldots \otimes |\alpha_L\rangle\,.
\label{eq:notation_2}
\ee
The coefficients $\kappa_{ij}$ have to be such that the matrix \eqref{eq:k_s} has a non-vanishing determinant. It was shown in \cite{PVCP18} that the product states with the building block \eqref{eq:integrable_block} are integrable according to the definition of \cite{PiPV17_2}. Furthermore, it was argued that, up to global $SU(3)$ transformations, one can always restrict themselves to the simpler case of diagonal boundary conditions

\bea 
| \psi_0 \rangle &=& \kappa_{11} |1,1\rangle +\kappa_{22} |2,2\rangle +  \kappa_{33} |3,3\rangle\,,
\label{eq:integrable_block_diagonal}
\eea
with
\be
|\kappa_{11}|\geq|\kappa_{22}|\geq |\kappa_{33}|\,,
\label{eq:maximal_magnetization}
\ee
which includes as a particular case the following $SO(3)$-invariant ``delta-state''
\be
| \psi_\delta \rangle = \frac{1}{\sqrt{3}}\left(|1,1\rangle + |2,2\rangle + |3,3\rangle \right)\,.
\label{eq:deltastate}
\ee

As  explained in \cite{PVCP18}, the relation between integrable states and transfer matrices with soliton-non-preserving boundary conditions can be derived by studying the following partition function
\be
\mathcal{Z}(\beta)=\langle \Psi_0|e^{-\beta H_L}|\Psi_0\rangle\,.
\label{eq:euclidean_partition_function}
\ee
Indeed, using the following Suzuki-Trotter decomposition~\cite{Klum92,FuKl99}
\be
e^{-\beta H_L}=\left[\bar{t}\left(-\frac{i\beta}{N}\right)t\left(-\frac{i\beta}{N}\right)\right]^{N/2}\,,
\ee
one can show that, for a given integrable initial state with two-site block \eqref{eq:integrable_block}, the partition function \eqref{eq:euclidean_partition_function} can be expressed as
\be
\mathcal{Z}(\beta)=\lim_{N\to\infty}\tr\left[\mathcal T_N^{L/2}\right]
\label{eq:key_relation0}
\ee
where
\be 
\mathcal{T}_N =\frac{1}{\langle \psi_0|\psi_0\rangle(1-\beta/2N)^{2N}}\tau^s_N(0)\,.
\label{eq:key_relation}
\ee 
Here $\tau^s_N(\lambda)$ is a soliton-non-preserving boundary transfer matrix acting on a chain of $N$ sites, which is related to \eqref{eq:def_tau} by a shift in the spectral parameters and rapidities
\be 
\lambda \to \lambda -3i/4\,, \qquad \xi_i \to \xi_i^s =\xi_i +3i/4\,. 
\label{eq:shift_parameters}
\ee 
Namely
\be
\tau^s_N(\lambda)={\rm tr}_a\left\{K^{+}T^s_a(\lambda)K^{-}\hat{T}^s_{\bar{a}}(\lambda)\right\}\,,
\label{eq:open_transfer_matrix}
\ee
where 
\bea
\fl T^s_a(\lambda)=R_{aN}(\lambda-\xi^s_N)\bar{R}_{a(N-1)}(\lambda-\xi^s_{N-1})\ldots R_{a2}(\lambda-\xi^s_2) \bar{R}_{a1}(\lambda-\xi^s_1)\,,\label{eq:shifted_monodromy}\\
\fl \hat{T}^s_{\bar{a}}(\lambda)=R_{1a}(\lambda+\xi^s_1-3i/2)\bar{R}_{2a}(\lambda+\xi^s_2-3i/2)\ldots 
\nonumber\\
R_{(N-1)a}(\lambda+\xi^s_{N-1}-3i/2) \bar{R}_{Na}(\lambda+\xi^s_N-3i/2)\,,
\eea
with inhomogeneities
\bea 
\xi^s_{2i} &=&   \frac{i\beta}{N}\,, \label{eq:inhomogeneities_1} \\
\xi^s_{2i+1} &=&    -\frac{i\beta}{N} + 3i/2\,.
\label{eq:inhomogeneities_2}
\eea 
Note that with the same shift \eqref{eq:shift_parameters}, the transfer \eqref{eq:def_bar_tau} is rewritten as
\bea
\bar{\tau}^s_N(\lambda)&=&{\rm tr}_a\left\{K_{\bar{a}}^{+}T^s_{\bar{a}}(\lambda)K_{\bar{a}}^{-}\hat{T}^s_a(\lambda)\right\}\,,\label{eq:bar_open_transfer_matrix}
\eea
where now
\bea
\fl T^s_{\bar{a}}(\lambda)=\bar{R}_{aN}(\lambda-\xi^s_{N})R_{a(N-1)}(\lambda-\xi^s_{N-1})\ldots \bar{R}_{a2}(\lambda-\xi^s_2) R_{a1}(\lambda-\xi^s_1)\,,\label{eq:shifted_monodromy_bar}\\
\fl \hat{T}^s_{a}(\lambda)=\bar{R}_{1a}(\lambda+\xi^s_1-3i/2)R_{2a}(\lambda+\xi^s_2-3i/2)\ldots \nonumber\\
\bar{R}_{(N-1)a}(\lambda+\xi^s_{N-1}-3i/2) R_{Na}(\lambda+\xi^s_{N}-3i/2)\,.
\eea
The $K$-matrices $K^\pm$ in \eqref{eq:open_transfer_matrix} and \eqref{eq:bar_open_transfer_matrix} are given by $K^-=K$,  $K^+=K^t$, where $K$ is defined in \eqref{eq:K_symmetric}.

Eqs.~\eqref{eq:key_relation0} and \eqref{eq:key_relation} provide the bridge between quantum quenches from integrable states and soliton-non-preserving boundary transfer matrices. In fact, as first observed in \cite{PiPV17}, several properties of the quench can be deduced from the study of such transfer matrices. Most prominently, we will show in Sec.~\ref{sec:lochmidt} how the leading eigenvalue of $\tau^s_N(\lambda)$ directly yields the Loschmidt echo after the quench, which is closely related to the partition function $\mathcal{Z}(\beta)$. In the next section we will derive a very important set of functional relations for the eigenvalues of the transfer matrix $\tau^s_N(\lambda)$. On the one hand,  these can be used to compute the spectrum of the latter; on the other hand, as we showed in Ref.~\cite{PVCP18}, they can also be used to derive the quasi-particle distribution functions of the steady state reached at long times.

\section{Fusion relations of boundary transfer matrices}\label{sec:fusion_relations}

The computation of the spectrum of $\tau^s_N(\lambda)$ for generic boundary conditions is a non-trivial problem. In fact, even for the well-known case of $XXZ$ Heisenberg chains, this has been solved only recently for finite system sizes \cite{nepomechie-02,clsw-03,fgsw-11,niccoli-12,cysw-13,nepomechie-13,Nepo}. Furthermore, non-trivial challenges arise for the study of the large-$N$ limit. A convenient strategy to address the computation of the spectrum of generic boundary transfer matrices, both for finite and infinite system sizes, was pursued in \cite{PiPV17}, by exploiting a set of functional relations which go under the name of $T$-system \cite{KuNS11} and which is obtained by an appropriate ``fusion procedure'' \cite{KuRS81}. In practice, one exploits the fact that the boundary transfer matrix can be embedded into a family of commuting operators, such that functional equations between their eigenvalues can be established, and eventually solved. 

In the case of $XXZ$ spin-$1/2$ chains, the $T$-system for boundary transfer matrices was worked out in \cite{Zhou95,Zhou96}, while to our knowledge it has not been explicitly obtained yet for $SU(3)$-invariant soliton-non-preserving boundary transfer matrices. This task is carried out in this section.  Applications of this result will be presented in the subsequent sections.

\subsection{$T$- and $Y$-systems for periodic transfer matrices}

It is useful to start our discussion from the case of periodic transfer matrices. It is an established result \cite{BaRe90,KuNS94} that the operators $t(\lambda)$ defined in \eqref{eq:periodic_transfer_matrix} can be embedded into a family of commuting transfer matrices $\{t_{m}^{(a)}(\lambda)\}$, with $m=1,2,\ldots +\infty$ and $a=1,2$, such that
\be
[t^{(a)}_m(\lambda),t^{(b)}_n(\mu)]=0\,.
\ee
These satisfy a set of functional relations which go under the name of $T$-system~\cite{Tsub03}
\bea
t^{(1)}_m\left(u+\frac{i}{2}\right)t^{(1)}_m\left(u-\frac{i}{2}\right)&=&t^{(1)}_{m+1}(u)t^{(1)}_{m-1}(u)+\Phi^{(1)}_{m}(u)t^{(2)}_{m}(u)\,, \label{eq:su3_t_system1} \\ 
t^{(2)}_m\left(u+\frac{i}{2}\right)t^{(2)}_m\left(u-\frac{i}{2}\right)&=&t^{(2)}_{m+1}(u)t^{(2)}_{m-1}(u)+\Phi^{(2)}_{m}(u)t^{(1)}_{m}(u)\,,
\label{eq:su3_t_system2}
\eea
with the convention
\bea
t_0^{(1)}(u)&\equiv& t_0^{(2)}(u)\equiv 1\,,\\
t_1^{(1)}(u)&=&t(u)\,,\\
t_1^{(2)}(u)&=&\bar{t}(u)\,,
\eea
and where $\bar{t}(\lambda)$ is defined in \eqref{eq:periodic_bar_transfer_matrix}. The operators $t^{(a)}_m(\lambda)$ can be explicitly obtained from $t(\lambda)$ and $\bar{t}(\lambda)$ by means of a geometrical construction called fusion \cite{KuRS81}. For this reason, the operators $t^{(a)}_m(\lambda)$ are called fused transfer matrices.

Each transfer matrix $t^{(a)}_m(\lambda)$, can be written as \eqref{eq:periodic_transfer_matrix}, but with a different irreducible representation of $SU(3)$ in the auxiliary space. We recall that each irreducible representation is labeled by a pair $(m_1,m_2)$ of two non-negative integers \cite{Slan81}. The $m$-fold symmetric tensor of the fundamental representation is associated with $(m,0)$, while the $m$-fold symmetric tensor of its conjugate representation is associated with the pair $(0,m)$. The dimension of a representation corresponding to the pair $(m_1,m_2)$ is given by the formula \cite{Slan81}
\be
d(m_1,m_2)=\frac{1}{2}(m_1+1)(m_2+1)(m_1+m_2+2)\,.
\ee
For each irreducible representation $(n,m)$, we label as $t_{(n,m)}(\lambda)$ the transfer matrix which has that representation in the auxiliary space. Note that, in the $SU(2)$ case, irreducible representations are instead labeled by a single index, and so all the possible transfer matrices are labeled by a single index. The operators $t^{(a)}_m(\lambda)$ entering the $T$-system are identified as follows
\bea
t^{(1)}_{j}(u)&=&t_{(j,0)}(u)\,,\\
t^{(2)}_{j}(u)&=&t_{(0,j)}(u)\,.
\eea

The validity of the $T$-system \eqref{eq:su3_t_system1}, \eqref{eq:su3_t_system2} was proven in \cite{KuNS94}. The functions $\Phi_n^{(1)}(u)$, $\Phi_n^{(2)}(u)$ are scalar and depend on the physical situation. In the case of thermal quantum transfer matrices, their explicit form can be found in \cite{Tsub03}. In any case, they can be seen to satisfy the relations
\bea
\Phi^{(1)}_{m}\left(u+\frac{i}{2}\right)\Phi^{(1)}_{m}\left(u-\frac{i}{2}\right)&=&\Phi^{(1)}_{m+1}\left(u\right)\Phi^{(1)}_{m-1}\left(u\right)\,,\label{eq:phi_relation1}\\
\Phi^{(2)}_{m}\left(u+\frac{i}{2}\right)\Phi^{(2)}_{m}\left(u-\frac{i}{2}\right)&=&\Phi^{(2)}_{m+1}\left(u\right)\Phi^{(2)}_{m-1}\left(u\right)\,,\label{eq:phi_relation2}
\eea
with
\bea
\Phi^{(1)}_0(u)=\Phi^{(2)}_0(u)=1\,.
\eea

From the $T$-system encoded in \eqref{eq:su3_t_system1} and \eqref{eq:su3_t_system2}, it is possible to derive another crucial set of functional relations which is usually called the $Y$-system \cite{KuNS11}. The latter is important in particular to obtain a description in terms of integral equations for functions defined on the real line. Following e.g. \cite{Tsub03} we define the $Y$-functions
\bea
y^{(1)}_m(u)&=&\frac{t^{(1)}_{m+1}(u)t^{(1)}_{m-1}(u)}{\Phi^{(1)}_m(u)t^{(2)}_{m}(u)}\,,\label{eq:def_y_function1}\\
y^{(2)}_m(u)&=&\frac{t^{(2)}_{m+1}(u)t^{(2)}_{m-1}(u)}{\Phi^{(2)}_m(u)t^{(1)}_{m}(u)}\,,\label{eq:def_y_function2}
\eea
which are easily seen to satisfy the $Y$-system
\bea
y^{(1)}_j\left(u+\frac{i}{2}\right)y^{(1)}_j\left(u-\frac{i}{2}\right)&=&\frac{\left[1+y^{(1)}_{j-1}(u)\right]\left[1+y^{(1)}_{j+1}(u)\right]}{1+\left[y^{(2)}_{j}(u)\right]^{-1}}\,,\label{eq:y_system1}\\
y^{(2)}_j\left(u+\frac{i}{2}\right)y^{(2)}_j\left(u-\frac{i}{2}\right)&=&\frac{\left[1+y^{(2)}_{j-1}(u)\right]\left[1+y^{(2)}_{j+1}(u)\right]}{1+\left[y^{(1)}_{j}(u)\right]^{-1}}\,,\label{eq:y_system2}
\eea
with the convention
\be
y^{(a)}_0(u)\equiv 0\,.
\ee
In the next sections, we will see how to transform these kinds of functional relations into integral equations.

\subsection{The $T$-system for open transfer matrices}

Analogously to the periodic case, also transfer matrices with open boundary conditions can be embedded into a family of fused operators $\tau_{(m_1,m_2)}(\lambda)$, characterized by a $SU(3)$ representation in the auxiliary space, corresponding to the pair $(m_1,m_2)$. As in the previous section we define
\bea
\tau^{(1)}_{j}(u)&=&\tau_{(j,0)}(u)\,,\\
\tau^{(2)}_{j}(u)&=&\tau_{(0,j)}(u)\,,
\eea
with the convention
\bea
\tau_0^{(1)}(u)&\equiv& \tau_0^{(2)}(u)\equiv 1\,,\\
\tau_1^{(1)}(u)&=&\tau(u)\,,\\
\tau_1^{(2)}(u)&=&\bar{\tau}(u)\,,
\eea
and where $\tau(\lambda)$ and $\bar{\tau}(\lambda)$ are defined in \eqref{eq:def_tau} and \eqref{eq:def_bar_tau}.

In the case of $XXZ$ spin-$1/2$ Heisenberg chains, fusion relations for open transfer matrices were derived in \cite{Zhou96}, where it was found that the latter satisfy the same $T$-system of periodic transfer matrices. A direct generalization of the work \cite{Zhou96} for the $SU(3)$-invariant spin chain is missing in the literature, even though some particular fused transfer matrices have been constructed, both for the soliton-preserving and soliton-non-preserving cases. For example, in the paper \cite{AACD04} the transfer matrix $\tau_{(1,1)}(u)$ was constructed in the soliton-non-preserving case from fusion of $\tau_{(1,0)}(u)$ and $\tau_{(0,1)}(u)$.

The purpose of this section is to write down the $T$-system in the case of boundary transfer matrices in the soliton-non-preserving case. In analogy with the spin-$1/2$ case, we assume that the latter is of the same form of the periodic case encoded in \eqref{eq:su3_t_system1} and \eqref{eq:su3_t_system2}. Hence, we only need to determine the auxiliary functions $\Phi_n^{(a)}(\lambda)$. In fact, given the recursive relations \eqref{eq:phi_relation1} and \eqref{eq:phi_relation2}, the whole calculation boils down to determining the functions $\Phi_1^{(1)}(\lambda)$ and $\Phi_1^{(2)}(\lambda)$.  In order to obtain the latter, we compute directly the products $\tau(\lambda)\tau(\mu)$ and $\bar{\tau}(\lambda)\bar{\tau}(\mu)$. As we will see in the following, by means of the fusion of $R$- and $K$-matrices introduced in \cite{Doik00,AACD04}, one can explicitly see that these products decompose exactly as in the functional relations \eqref{eq:su3_t_system1} and \eqref{eq:su3_t_system2} for $n=1$.

In the following calculations we will restrict to the case of invertible $K$-matrices, which are solution to the reflection equations \eqref{eq:reflection_1} and \eqref{eq:reflection_2}. In particular, according to the results of \cite{AACD04}, we know that the latter are of the form \eqref{eq:K_symmetric}, which will be explicitly exploited.

\subsection{Fusion of $R$-matrices}

As a first ingredient, we need to review the fusion properties of the $R$-matrix \cite{KuRS81,AACD04}. We note that $R_{23}(-i)=P_{\mathcal{H}_3}$, where $P_{\mathcal{H}_3}$ is the projector onto the three-dimensional space in the decomposition
\be
3_{2}\otimes 3_{3}= 6_{23} \oplus \bar{3}_{23}\,,
\label{eq:decomposition}
\ee
where we denote with $d$ and $\bar{d}$ the irreducible fundamental and conjugate representation of dimension $d$ respectively. Consider now the Yang-Baxter equations \eqref{eq:YBER11}, \eqref{eq:YBER12}. Choosing $\mu=-i$ it is immediate to see from \eqref{eq:YBER12} that the operators
\bea
R_{13}(\lambda-i)R_{12}(\lambda)\,,\qquad 
\bar{R}_{13}(\lambda-i)\bar{R}_{12}(\lambda)\,,
\label{eq:product1}
\eea
leave stable the symmetric six-dimensional representation in the decomposition \eqref{eq:decomposition}, which implies
\bea
R_{13}(\lambda-i)R_{12}(\lambda)P_{23}^{+}&=&P_{23}^{+}R_{13}(\lambda-i)R_{12}(\lambda)P_{23}^{+}\,,\label{eq:id_1}\\
P_{23}^{-}R_{13}(\lambda-i)R_{12}(\lambda)&=&P_{23}^{-}R_{13}(\lambda-i)R_{12}(\lambda)P_{23}^{-}\,,\label{eq:id_2}
\eea
where $P_{ab}^{-}=P_{\mathcal{H}_3}$ while $P_{ab}^{+}$ is the projector on the six-dimensional symmetric representation in the decomposition \eqref{eq:decomposition}. Analogously, one can see that the operators
\bea
R_{12}(\lambda)R_{13}(\lambda-i)\,,\qquad 
\bar{R}_{12}(\lambda)\bar{R}_{13}(\lambda-i)\,,
\label{eq:product2}
\eea
leave stable the anti-symmetric three-dimensional representation in the decomposition \eqref{eq:decomposition}. Accordingly, one can define the fused $R$-matrices
\bea
R_{1,\langle 23\rangle }^{(2)}(\lambda)&=&P_{23}^{+}R_{13}(\lambda-i)R_{12}(\lambda)P_{23}^{+}\,,\label{eq:fused_R1}\\
\bar{R}_{1,\langle 23\rangle }^{(2)}(\lambda)&=&P_{23}^{+}\bar{R}_{13}(\lambda-i)\bar{R}_{12}(\lambda)P_{23}^{+}\,,\label{eq:fused_R2}
\eea
where two indices between brackets stand for one single index labeling the six-dimensional representation. These fused $R$-matrices also satisfy the Yang-Baxter relations, as it can be easily seen for example through a graphical proof.

\subsection{Fusion of $K$-matrices}

Analogously to the case of $R$-matrices, one can also address the fusion of boundary reflection matrices \cite{MeNe92}.
We consider the soliton-non-preserving case, and start with the reflection equation \eqref{eq:reflection_1}. Setting $\mu=\lambda+i$ we find
\be
P_{ab}^-K_a(\lambda)\bar{R}_{ba}(2\lambda+i)K_b(\lambda+i)=K_b(\lambda+i)\bar{R}_{ab}(2\lambda+i)K_a(\lambda)P_{ab}^-\,,
\ee
from which we obtain straightforwardly that the operator
\be
K_a(\lambda)\bar{R}_{ba}(2\lambda+i)K_b(\lambda+i)
\label{eq:almost_fused_sym}
\ee
leaves stable the symmetric six-dimensional representation in the decomposition \eqref{eq:decomposition}. Analogously, the operator
\be
K_b(\lambda+i)\bar{R}_{ab}(2\lambda+i)K_a(\lambda)
\ee
leaves stable the anti-symmetric three dimensional representation in the decomposition \eqref{eq:decomposition}. Note that in the definition of the boundary transfer matrix, one has the matrices
\bea
K_a^{-}(\lambda)=K_a(\lambda)\,,\qquad 
K_a^{+}(\lambda)=K_a^{t}(-\lambda-3i/2)\,.
\eea
Accordingly, using the considerations above, one can see that
\bea
\fl K^{-}_a(\lambda)\bar{R}_{ba}(2\lambda+i)K^{-}_b(\lambda+i)P^{+}_{ab}=P^{+}_{ab}K^{-}_a(\lambda)\bar{R}_{ba}(2\lambda+i)K^{-}_b(\lambda+i)P^{+}_{ab}\,,\\
\fl P^{-}_{ab}K^{-}_a(\lambda)\bar{R}_{ba}(2\lambda+i)K^{-}_b(\lambda+i)=P^{-}_{ab}K^{-}_a(\lambda)\bar{R}_{ba}(2\lambda+i)K^{-}_b(\lambda+i)P^{-}_{ab}\,,\\
\fl P^{-}_{ab}K^{+}_b(\lambda+i)\bar{R}_{ab}(2\lambda+i)K^{+}_a(\lambda)=P^{-}_{ab}K^{+}_b(\lambda+i)\bar{R}_{ab}(2\lambda+i)K^{+}_a(\lambda)P_{ab}^-\,,\\
\fl K^{+}_b(\lambda+i)\bar{R}_{ab}(2\lambda+i)K^{+}_a(\lambda)P^{+}_{ab}=P^{+}_{ab}K^{+}_b(\lambda+i)\bar{R}_{ab}(2\lambda+i)K^{+}_a(\lambda)P_{ab}^+\,.
\eea
Hence, as in the previous section, one can define the fused $K$-matrices
\bea
K_{\langle ab\rangle }^{(2),-}(\lambda,\mu)&=&P_{ab}^{+}K_b^{-}(\lambda)\bar{R}_{ab}(2\lambda+i)K_a^{-}(\lambda  +i)P_{ab}^{+}\,,\label{eq:fused_K1}\\
K_{\langle ab\rangle}^{(2),+}(\lambda,\mu)&=&P_{ab}^{+}K_a^{+}(\lambda+i)\bar{R}_{ab}(2\lambda+i)K_b^{+}(\lambda)P_{ab}^{+}\,,\label{eq:fused_K2}
\eea
which are seen, once again, to satisfy an appropriate fused version of the Yang-Baxter reflection equations.

\subsection{Product of transfer matrices}

We can now make a step forward and compute the product of two transfer matrices $\tau(\lambda)$, as defined in \eqref{eq:def_tau}. We start with the identity
\bea
\fl \tau(\mu)\tau(\lambda)={\rm tr}_{b}\left\{K_b^{+}(\mu)T_b(\mu)K_b^{-}(\mu)\hat{T}_{\bar{b}}(\mu)\right\}{\rm tr}_{a}\left\{T^{t_a}_a(\lambda)\left[K_a^{+}(\lambda)\right]^{t_a}\hat{T}^{t_a}_{\bar{a}}(\lambda)\left[K_a^{-}(\lambda)\right]^{t_a}\right\}\nonumber\\
\fl =\frac{1}{\zeta(\lambda+\mu+3i/2)}{\rm tr}_{a\otimes b}\left\{T_b(\mu)K_b^{-}(\mu)\left[K_a^{-}(\lambda)\right]^{t_a}\right.\nonumber\\
\fl \times \left.\bar{R}^{t_a}_{ab}(-\mu-\lambda-3i) T^{t_a}_a(\lambda)\hat{T}_{\bar{b}}(\mu)\bar{R}^{t_a}_{ab}(\mu+\lambda)\left[K_a^{+}(\lambda)\right]^{t_a}K_b^{+}(\mu)\hat{T}^{t_a}_{\bar{a}}(\lambda)\right\}\,,
\label{eq:intermediate_step}
\eea
where we used the crossing unitarity
\be
\bar{R}^{t_1}_{12}(\lambda)\bar{R}^{t_1}_{12}(-\lambda-3i)=\zeta(\lambda+3i/2)\,.
\ee
We now make use of the following property of the partial transpose. Let $M$ be an operator acting on the product $\mathcal{H}_a\otimes \mathcal{H}_b\otimes \mathcal{H}_p$, and $\mathcal{B}$ an operator acting on $\mathcal{H}_b\otimes \mathcal{H}_p$. Then
\bea
\left[M\mathcal{B}\right]^{t_a}&=&M^{t_a}\mathcal{B}\,.
\label{eq:identity_trace}
\eea
Applying \eqref{eq:identity_trace} to Eq.~\eqref{eq:intermediate_step} we get
\bea
\fl \tau(\mu)\tau(\lambda)=\frac{1}{\zeta(\lambda+\mu+3i/2)}{\rm tr}_{a\otimes b}\left\{T_b(\mu)\left[K_b^{-}(\mu)\bar{R}_{ab}(-\mu-\lambda-3i)K_a^{-}(\lambda)\right]^{t_a}T^{t_a}_a(\lambda)\right.\nonumber\\
\times \left. \hat{T}_{\bar{b}}(\mu)\left[K_a^{+}(\lambda)\bar{R}_{ab}(\mu+\lambda)K_b^{+}(\mu)\right]^{t_a}\hat{T}^{t_a}_{\bar{a}}(\lambda)\right\}\,.
\label{eq:intermediate_step2}
\eea
Next, define
\bea
\mathcal{K}_{ab}^{-}(\lambda,\mu)&=&K_b^{-}(\mu)\bar{R}_{ab}(-\mu-\lambda-3i)K_a^{-}(\lambda)\,,\\
\mathcal{K}_{ab}^{+}(\lambda,\mu)&=&K_a^{+}(\lambda)\bar{R}_{ab}(\mu+\lambda)K_b^{+}(\mu)\,,
\eea
so that
\bea
\fl \tau(\mu)\tau(\lambda)=\frac{1}{\zeta(\lambda+\mu+3i/2)}{\rm tr}_{a\otimes b}\left\{T_b(\mu)\left[\mathcal{K}_{ab}^{-}\right]^{t_a}T^{t_a}_a(\lambda) \hat{T}_{\bar{b}}(\mu)\left[\mathcal{K}_{ab}^{+}\right]^{t_a}\hat{T}^{t_a}_{\bar{a}}(\lambda)\right\}\,,
\label{eq:intermediate_step3}
\eea
where we omitted the dependence of $\mathcal{K}_{ab}^{\pm}$ on the spectral parameters. Finally, we compute
\bea
 {\rm tr}_{a\otimes b}\left\{T_b(\mu)\left[\mathcal{K}_{ab}^{-}\right]^{t_a}T^{t_a}_a(\lambda) \hat{T}_{\bar{b}}(\mu)\left[\mathcal{K}_{ab}^{+}\right]^{t_a}\hat{T}^{t_a}_{\bar{a}}(\lambda)\right\}
\nonumber\\
 =\left\{T_b(\mu)\left[\mathcal{K}_{ab}^{-}\right]^{t_a}T^{t_a}_a(\lambda) \hat{T}_{\bar{b}}(\mu)\left[\mathcal{K}_{ab}^{+}\right]^{t_a}\hat{T}^{t_a}_{\bar{a}}(\lambda)\right\}_{\alpha,\beta,\varphi}^{\alpha,\beta,\varphi}\nonumber\\
 =\left[T_b(\mu)T_a(\lambda)\mathcal{K}_{ab}^{-}\right]_{a_2,\beta,\varphi}^{\alpha,b_2,p_2}\left[\hat{T}_{\bar{b}}(\mu)\hat{T}_{\bar{a}}(\lambda)\mathcal{K}_{ab}^{+}\right]^{a_2,\beta,\varphi}_{\alpha,b_2,p_2}\nonumber\\
={\rm tr}_{a\otimes b}\left\{T_b(\mu)T_a(\lambda)\mathcal{K}_{ab}^{-}\hat{T}_{\bar{b}}(\mu)\hat{T}_{\bar{a}}(\lambda)\mathcal{K}_{ab}^{+}\right\}\,,
\eea
where in the second and third line we have introduced contracted indices $\alpha, \beta, \varphi, a_2, b_2, p_2$ spanning the basis of the auxilliary space. The final result of this computation is extremely simple and reads
\bea
\fl \tau(\mu)\tau(\lambda)=\frac{1}{\zeta(\lambda+\mu+3i/2)}{\rm tr}_{a\otimes b}\left\{T_b(\mu)T_a(\lambda)\mathcal{K}_{ab}^{-}(\lambda,\mu)\hat{T}_{\bar{b}}(\mu)\hat{T}_{\bar{a}}(\lambda)\mathcal{K}_{ab}^{+}(\lambda,\mu)\right\}\,.
\label{eq:final_formula}
\eea

As a second step of the derivation, we show that the product \eqref{eq:final_formula} decomposes into the sum of two terms, provided that $\lambda$ and $\mu$ are chosen appropriately. Our goal is to extract the function $\Phi^{(1)}_1(\lambda)$ which is needed to write down the $T$-system for general $n$. First, we define the fused transfer matrices $\tau^{(1)}_{2}(\lambda)$ and $\tau^{(2)}_{2}(\lambda)$ as
\bea
\tau^{(1)}_2(\lambda)&=&{\rm tr}_{a}\left\{K_a^{(2),+}(\lambda)T^{(2)}_a(\lambda)K_a^{(2),-}(\lambda)\hat{T}^{(2)}_{\bar{a}}(\lambda)\right\}\,,\label{eq:def_tau2}\\
\tau^{(2)}_2(\lambda)&=&{\rm tr}_a\left\{K_{\bar{a}}^{(2),+}(\lambda)T^{(2)}_{\bar{a}}(\lambda)K_{\bar{a}}^{(2),-}(\lambda)\hat{T}^{(2)}_a(\lambda)\right\}\,,\label{eq:def_bar_tau2}
\eea
where
\bea
\fl T^{(2)}_a(\lambda)=R_{aN}^{(2)}(\lambda-\xi_{N})\bar{R}^{(2)}_{a(N-1)}(\lambda-\xi_{N-1})\ldots R^{(2)}_{a2}(\lambda-\xi_2) \bar{R}^{(2)}_{a1}(\lambda-\xi_1)\,,\\
\fl \hat{T}^{(2)}_{\bar{a}}(\lambda)=R^{(2)}_{1a}(\lambda+\xi_1)\bar{R}^{(2)}_{2a}(\lambda+\xi_2)\ldots R^{(2)}_{(N-1)a}(\lambda+\xi_{N-1}) \bar{R}^{(2)}_{Na}(\lambda+\xi_{N})\,,\\
\fl T^{(2)}_{\bar{a}}(\lambda)=\bar{R}^{(2)}_{aN}(\lambda-\xi_{N})R^{(2)}_{a(N-1)}(\lambda-\xi_{N-1})\ldots \bar{R}^{(2)}_{a2}(\lambda-\xi_2) R^{(2)}_{a1}(\lambda-\xi_1)\,,\\
\fl \hat{T}^{(2)}_{a}(\lambda)=\bar{R}^{(2)}_{1a}(\lambda+\xi_1)R^{(2)}_{2a}(\lambda+\xi_2)\ldots \bar{R}^{(2)}_{(N-1)a}(\lambda+\xi_{N-1}) R^{(2)}_{Na}(\lambda+\xi_{N})\,.
\eea
Here we used the fused $R$- and $K$-matrices defined \eqref{eq:fused_R1},\eqref{eq:fused_R2} and \eqref{eq:fused_K1}, \eqref{eq:fused_K2} respectively. Note that the traces in \eqref{eq:def_tau2} and \eqref{eq:def_bar_tau2} are now over a $d$-dimensional space with $d=6$.

Next, we need to specify the $K$-matrices $K^{\pm}_{a}$ and $K^{\pm}_{\bar{a}}$ entering the definition of the boundary transfer matrices involved in the $T$-system which we intend to derive. First, we restrict to invertible solutions to the reflection equations \eqref{eq:reflection_1} and \eqref{eq:reflection_2}. Next, in order to construct a good set of fused transfer matrices, we need to start with two operators $\tau(\lambda)$ and $\bar{\tau}(\mu)$ such that $[\tau(\lambda),\bar{\tau}(\mu)]=0$. This enforces a relation between the $K$-matrices $K^{\pm}_{a}$ and $K^{\pm}_{\bar{a}}$ of $\tau(\lambda)$ and $\bar{\tau}(\mu)$ respectively. In fact, it was shown in \cite{AACD04}  that, in order for the transfer matrices $\tau(\lambda)$ and $\bar{\tau}(\mu)$ to commute, the $K$-matrix $K_{\bar{a}}$ has to be related to $K_a$ as
\be
K_{\bar{a}}= \alpha K_{a}^{-1}\,,
\label{eq:choice_k_bar}
\ee
where $\alpha$ is a non-vanishing constant number. In the following, we will choose $\alpha=1$, and derive the corresponding $T$-system. Explicitly, we will consider boundary transfer matrices $\tau(\lambda)$ and $\bar{\tau(\lambda)}$ as in \eqref{eq:def_tau} and \eqref{eq:def_bar_tau} with the $K$-matrices
\bea
K_{a}^{-}(\lambda)&\equiv &VK_s\,,\\
K_{a}^{+}(\lambda)&=&K_a^{t}(-\lambda-i3/2)\equiv V K_s\,,\\
K_{\bar{a}}^{-}(\lambda)&\equiv &K^{-1}_sV\,,\\
K_{a}^{+}(\lambda)&=&K_{\bar{a}}^{t}(-\lambda-i3/2)\equiv K^{-1}_sV\,,
\eea
where $K_{s}$ is any invertible symmetric numerical matrix. We can now plug the resolution of the identity 
\be
\mathbf{1}_{ab}=P_{ab}^{+}+P_{ab}^{-}\,.
\label{eq:resolution_identity}
\ee
into \eqref{eq:final_formula}, and setting $\mu=\nu-i/2$, $\lambda=\nu+i/2$ we obtain
\bea
\fl \tau(\nu-i/2)\tau(\nu+i/2)\nonumber\\
=\frac{1}{\zeta(2\nu+3i/2)} \left[{\rm tr}_{a\otimes b}\left\{T_b(\mu)T_a(\lambda)P_{ab}^{+}\mathcal{K}_{ab}^{-}(\lambda,\mu)\hat{T}_{\bar{b}}(\mu)\hat{T}_{\bar{a}}(\lambda)\mathcal{K}_{ab}^{+}(\lambda,\mu)\right\}\right.\nonumber\\
+\left.{\rm tr}_{a\otimes b}\left\{T_b(\mu)T_a(\lambda)P_{ab}^{-}\mathcal{K}_{ab}^{-}(\lambda,\mu)\hat{T}_{\bar{b}}(\mu)\hat{T}_{\bar{a}}(\lambda)\mathcal{K}_{ab}^{+}(\lambda,\mu)\right\}\right]\big|_{\mu,\lambda=\nu\mp i/2}\,.
\eea
From this expression one can see that the product $\tau_{b}(\nu-i/2)\tau_{a}(\nu+i/2)$ decomposes into the sum of two terms. In particular, it is straightforward to see
\be
\fl {\rm tr}_{a\otimes b}\left\{T_b(\mu)T_a(\lambda)P_{ab}^{+}\mathcal{K}_{ab}^{-}(\lambda,\mu)\hat{T}_{\bar{b}}(\mu)\hat{T}_{\bar{a}}(\lambda)\mathcal{K}_{ab}^{+}(\lambda,\mu)\right\}\big|_{\mu,\lambda=\nu\mp i/2}\propto \tau_{2}^{(1)}(\nu)\,.
\ee
In order to conclude our calculation, we show
\be
\fl {\rm tr}_{a\otimes b}\left\{T_b(\mu)T_a(\lambda)P_{ab}^{-}\mathcal{K}_{ab}^{-}(\lambda,\mu)\hat{T}_{\bar{b}}(\mu)\hat{T}_{\bar{a}}(\lambda)\mathcal{K}_{ab}^{+}(\lambda,\mu)\right\}\big|_{\mu,\lambda=\nu\mp i/2}\propto 	\tau_{1}^{(2)}(\nu)\,.
\ee
Making use of \eqref{eq:id_1} and \eqref{eq:id_2}, we get immediately
\bea
{\rm tr}_{a\otimes b}\left\{T_b(\mu)T_a(\lambda)P_{ab}^{-}\mathcal{K}_{ab}^{-}(\lambda,\mu)\hat{T}_{\bar{b}}(\mu)\hat{T}_{\bar{a}}(\lambda)\mathcal{K}_{ab}^{+}(\lambda,\mu)\right\}\big|_{\mu,\lambda=\nu\mp i/2}\nonumber\\
={\rm tr}_{a}\left\{\mathcal{T}_a(\nu)\mathcal{Q}_{\langle ab \rangle}^{-}(\nu)\hat{\mathcal{T}}_{\bar{a}}(\nu)\mathcal{Q}_{\langle ab \rangle}^{+}(\nu)\right\}\,,
\eea
where
\bea
\mathcal{T}_a(\lambda)=\mathcal{L}_{aN}(\lambda-\xi_{N})\bar{\mathcal{L}}_{a(N-1)}(\lambda-\xi_{N-1})\ldots \mathcal{L}_{a2}(\lambda-\xi_2) \bar{\mathcal{L}}_{a1}(\lambda-\xi_1)\,,\\
\hat{\mathcal{T}}_{\bar{a}}(\lambda)=\mathcal{L}_{1a}(\lambda+\xi_1)\bar{\mathcal{L}}_{2a}(\lambda+\xi_2)\ldots \mathcal{L}_{(N-1)a}(\lambda+\xi_{N-1}) \bar{\mathcal{L}}_{Na}(\lambda+\xi_{N})\,,
\eea
and
\bea
\mathcal{L}_{\langle ab\rangle,1}(\lambda)&=&P_{ab}^-R_{1a}(\lambda-i/2)R_{1b}(\lambda+i/2)P_{ab}^-\,,\\
\bar{\mathcal{L}}_{\langle ab\rangle,1}(\lambda)&=&P_{ab}^-\bar{R}_{1a}(\lambda-i/2)\bar{R}_{1b}(\lambda+i/2)P_{ab}^-\,,\\
\mathcal{Q}^-_{\langle ab\rangle }(\lambda)&=&P_{ab}^{-}K_b^{-}(\lambda-i/2)\bar{R}_{ab}(2\lambda)K_a^{-}(\lambda+i/2)P_{ab}^-\,,\\
\mathcal{Q}^+_{\langle ab\rangle }(\lambda)&=&P_{ab}^{-}K_a^{+}(\lambda+i/2)\bar{R}_{ab}(-2\lambda-3i)K_b^{+}(\lambda-i/2)P_{ab}^{-}\,.
\eea
One can straightforwardly compute
\bea
\mathcal{L}_{\langle ab\rangle, c}(\lambda)&=&(i/2-\lambda)W_{\langle ab\rangle}^{-1}\bar{R}^{(2)}_{\langle ab\rangle,c}(\lambda)W_{\langle ab\rangle}\,,\\
\bar{\mathcal{L}}_{\langle ab\rangle,c}(\lambda)&=&(2i+\lambda)W_{\langle ab\rangle}^{-1}R^{(2)}_{\langle ab\rangle,c}(\lambda)W_{\langle ab\rangle}\,,
\eea
where the matrices $R^{(2)}_{\langle ab\rangle,c}(\lambda)$ and $\bar{R}^{(2)}_{\langle ab\rangle,c}(\lambda)$ are defined in \eqref{eq:fused_R1}, \eqref{eq:fused_R2}, while
\be
W={\rm diag}(1,-1,1)\,.
\ee
Note that the following basis has been chosen for the fused spaced obtained out of $h_a{\otimes} h_a$
\bea
w_{1}&=&\frac{1}{\sqrt{2}}(-|2\rangle_{a}|3\rangle_{b}+|3\rangle_{a}|2\rangle_{b})\,,\\
w_{2}&=&\frac{1}{\sqrt{2}}(-|1\rangle_{a}|3\rangle_{b}+|3\rangle_{a}|1\rangle_{b})\,,\\
w_{3}&=&\frac{1}{\sqrt{2}}(-|1\rangle_{a}|2\rangle_{b}+|2\rangle_{a}|1\rangle_{b})\,.
\eea
Analogously, for any $K$-matrix which is constant, invertible and symmetric, one can compute
\bea
WP^{-}_{ab}K^{+}_a\bar{R}_{ab}(2\lambda)K^{+}_bP^{-}_{ab}W^{-1}=-\frac{(3i+4\lambda)}{2}[{\rm det}K_s]K_{s}^{-1}\,,\\
WP^{-}_{ab}K^{-}_a\bar{R}_{ab}(-2\lambda-3i)K^{-}_bP^{-}_{ab}W^{-1}=\frac{(3i+4\lambda)}{2}[{\rm det}K^{+}_s](K^{+}_{s})^{-1}\,.
\eea
Recalling Eq.~\eqref{eq:choice_k_bar} we can collect all the prefactors and obtain the final result of the calculation. Performing then analogous steps for the product $\bar{\tau}(\lambda-i/2)\bar{\tau}(\lambda+i/2)$ we get
\bea
\tau(\lambda-i/2)\tau(\lambda+i/2)&=&\tau_2^{(1)}(\lambda)+\Phi^{(1)}_1(\lambda)\bar{\tau}^{(2)}(\lambda)\,,\\
\bar{\tau}(\lambda-i/2)\bar{\tau}(\lambda+i/2)&=&\tau_2^{(2)}(\lambda)+\Phi^{(2)}_1(\lambda)\tau^{(2)}(\lambda)\,,
\eea
where
\bea
\Phi_1^{(1)}(\lambda)&=&-\frac{1}{4\zeta(2\lambda+3i/2)}\left[\prod_{j=1}^{N/2}\left(i/2-\lambda+\xi_{2j}\right)\left(2i+\lambda-\xi_{2j-1}\right)\right.\nonumber\\
&\times&\left.\left(i/2-\lambda-\xi_{2j-1}\right)\left(2i+\lambda+\xi_{2j}\right)\right.\Bigg](3i+4\lambda)^2\left[{\rm det}K_s\right]^2\,,\\
\label{eq:phi1}
\Phi_1^{(2)}(\lambda)&=&-\frac{1}{4\zeta(2\lambda+3i/2)}\left[\prod_{j=1}^{N/2}\left(2i+\lambda-\xi_{2j}\right)\left(i/2-\lambda+\xi_{2j-1}\right)\right.\nonumber\\
&\times&\left.\left(2i+\lambda+\xi_{2j-1}\right)\left(i/2-\lambda-\xi_{2j}\right)\right.\Bigg](3i+4\lambda)^2\left[{\rm det}K_s\right]^{-2}\,.
\label{eq:phi2}
\eea
From this result, we can finally exploit the functional relations \eqref{eq:phi_relation1} and \eqref{eq:phi_relation2} to obtain higher functions $\Phi^{(1)}_n(\lambda)$ and $\Phi^{(2)}_n(\lambda)$. 

\subsection{The final result}

We can now summarize the final result of our calculations, namely the explicit $T$-system for soliton-non-preserving open boundary conditions. It is convenient to write down the latter for the transfer matrices $\tau^s_N(\lambda)$ with shifted spectral parameters defined in \eqref{eq:open_transfer_matrix}. After performing the shift \eqref{eq:shift_parameters} in the formulas of the previous section, we find that the fused transfer matrices obtained from $\tau_N(\lambda)$ satisfy the $T$-system
\bea
\tau^{(1)}_m\left(u+\frac{i}{2}\right)\tau^{(1)}_m\left(u-\frac{i}{2}\right)&=&\tau^{(1)}_{m+1}(u)\tau^{(1)}_{m-1}(u)+	\tilde{\Phi}^{(1)}_{m}(u)\tau^{(2)}_{m}(u)\,, \label{eq:boundary_t_system1} \\ 
\tau^{(2)}_m\left(u+\frac{i}{2}\right)\tau^{(2)}_m\left(u-\frac{i}{2}\right)&=&\tau^{(2)}_{m+1}(u)\tau^{(2)}_{m-1}(u)+\tilde{\Phi}^{(2)}_{m}(u)\tau^{(1)}_{m}(u)\,,
\label{eq:boundary_t_system2}
\eea
with the convention
\bea
\tau_0^{(1)}(u)&\equiv& \tau_0^{(2)}(u)\equiv 1\,,\\
\tau_1^{(1)}(u)&=&\tau^s_N(u)\,,\\
\tau_1^{(2)}(u)&=&\bar{\tau}^s_N(u)\,,
\eea
and where $\tau^s_N(u)$ and $\bar{\tau}^s_N(u)$ are defined in \eqref{eq:open_transfer_matrix} and \eqref{eq:bar_open_transfer_matrix}. Here the functions $\tilde{\Phi}_{m}^{(a)}(\lambda)$ are given by
\bea
\tilde{\Phi}_n^{(1)}(\lambda)=\prod_{j=1}^{n}f^{(1)}\left[\lambda-(n-2j+1)\frac{i}{2}\right]\,,\label{eq:final_result_1}\\
\tilde{\Phi}_n^{(2)}(\lambda)=\prod_{j=1}^{n}f^{(2)}\left[\lambda-(n-2j+1)\frac{i}{2}\right]\,,\label{eq:final_result_2}
\eea
where
\bea
f^{(1)}(\lambda)&=&-\frac{4\lambda^2}{\zeta(2\lambda)}\left[\prod_{j=1}^{N/2}\left(i/2-\lambda+\xi^s_{2j}\right)\left(2i+\lambda-\xi^s_{2j-1}\right)\right.\nonumber\\
&\times&\left.\left(2i-\lambda-\xi^s_{2j-1}\right)\left(i/2+\lambda+\xi^s_{2j}\right)\right]\left[{\rm det}K_s\right]^2\,,\label{eq:final_result_3}\\
f^{(2)}(\lambda)&=&-\frac{4\lambda^2}{\zeta(2\lambda)}\left[\prod_{j=1}^{N/2}\left(i/2-\lambda+\xi^s_{2j-1}\right)\left(2i+\lambda-\xi^s_{2j}\right)\right.\nonumber\\
&\times&\left.\left(2i-\lambda-\xi^s_{2j}\right)\left(i/2+\lambda+\xi^s_{2j-1}\right)\right]\left[{\rm det}K_s\right]^{-2}\,.\label{eq:final_result_4}
\eea
Note that the general expressions \eqref{eq:final_result_1} and \eqref{eq:final_result_2} are obtained exploiting the relations \eqref{eq:phi_relation1} and \eqref{eq:phi_relation2}, and using the convention $\tilde{\Phi}^{(1)}_0\equiv \tilde{\Phi}^{(2)}_0\equiv 1$. This result generalizes the findings of \cite{Zhou96} to $SU(3)$-invariant boundary transfer matrices in the case of soliton-non-preserving boundary conditions, and will be used throughout the rest of this work.

\section{The $Y$-system and the Loschmidt echo}
\label{sec:Loschmidt}

As we have anticipated, the usefulness of the $T$-system in \eqref{eq:boundary_t_system1}, \eqref{eq:boundary_t_system2} relies on the possibility of solving it for any length $N$, and also in the limit $N\to\infty$. In turn, this provides an explicit solution to the problem of computing the spectrum of the boundary transfer matrix $\tau^s_N(\lambda)$. In practice, this is done as follows. First, one needs to write down the $Y$-system corresponding to the $T$-system; then the latter is cast, within a standard procedure, into a set of non-linear integral equations, which can be finally solved numerically. Since the $T$-system in the case of boundary transfer matrices is the same of the periodic case, also the $Y$-system \cite{KuNS11} will have the exact same form, encoded in Eqs.~\eqref{eq:y_system1} and \eqref{eq:y_system2}. Then, we are left with the task of explicitly writing down the corresponding non-linear integral equations. This procedure is well-known in the literature, and has been repeatedly used also in previous applications of the QTM formalism to quantum quenches \cite{PiPV17,PiPV18}. For this reason, here we only briefly summarize it, referring to the work \cite{PiPV17} for a detailed explanation of the method.

In summary, one needs to take the logarithmic derivative of both sides of \eqref{eq:y_system1} and \eqref{eq:y_system2}, and then perform a Fourier transform to obtain integrals along segments with constant imaginary parts. The integration contours can then be shifted back to the real axis, by carefully taking into account all the singularities of the logarithmic derivatives within the \emph{physical strip}, which is the portion of the complex plane defined as
\be
\mathcal{S}=\left\{\lambda\left|-\frac{1}{2}\leq{\rm Im}(\lambda)\leq \frac{1}{2}\right.\right\}\,,
\label{eq:physical_strip}
\ee
where ${\rm Im}(\lambda)$ denotes the imaginary part of the complex number $\lambda$. From the above discussion, it is obvious that the final form of the integral equations will essentially depend on the analytic structure of the $Y$-functions inside the physical strip. Different boundary matrices $K(\lambda)$ and different inhomogeneities $\xi_j$ will give rise to different singularities, and ultimately to different integral equations (in fact, differences will also arise in the asymptotic behavior of $y^{(r)}_n(\lambda)$ as $n\to\infty$, cf. Ref.~\cite{PiPV17}).

In order to illustrate the above procedure in detail, we consider the problem of computing the leading eigenvalue of the boundary transfer matrix associated to the delta-state \eqref{eq:deltastate} (with inhomogeneities \eqref{eq:inhomogeneities_1} and \eqref{eq:inhomogeneities_2}). As we will see in Sec.~\ref{sec:lochmidt}, this will immediately give us access to the Loschmidt echo at imaginary and real times after the quench from the delta-state. Once again, we stress that, even though we focus on a particular boundary transfer matrix corresponding to a given boundary state, our treatment is quite general, and analogous calculations can be carried out in more general cases.

\subsection{The non-linear integral equations for the delta-state}

The delta-state \eqref{eq:deltastate} is associated with the boundary transfer matrix \eqref{eq:open_transfer_matrix} with $K$-matrices $K^-V = K^+V = \mathbf{1}$. The inhomogeneties are chosen as \eqref{eq:inhomogeneities_1} and \eqref{eq:inhomogeneities_2}, which corresponds to imaginary time evolution. At the end of the calculation, we will address the real-time case.

First, we note that in this case the boundary transfer matrix commutes with the global spin operator 
\be 
S^z = \sum_{i=1}^{M}  \left(\begin{array}{ccc} 1&0&0\\0&0&0\\0&0&-1\\ \end{array} \right)_{i}\,,
\ee 
and that the leading eigenvalue of $\tau^s_N(0)$ lies in the $S^z=0$ sector. In order to get an intuition of the corresponding $Y$-functions $y^{(1)}_{1}(\lambda)$ and $y^{(2)}_{1}(\lambda)$, we compute it numerically in the physical strip \eqref{eq:physical_strip} for small values of $N$. Remarkably, this can be done even without the knowledge of the Bethe roots as follows. First, one computes the leading eigenvector of $\tau^s_N(0)$  using exact diagonalization calculations for a finite $N$. Next, since all of the transfer matrices introduced in the previous section commute with one another, one can act on the latter with $\tau^s_N(\lambda)$ and $\bar{\tau}^s_N(\lambda)$ to get the associated eigenvalues as a function of $\lambda$. Then, the $Y$-functions $y^{(1)}_{1}(\lambda)$ and $y^{(2)}_{1}(\lambda)$ are obtained straightforwardly from \eqref{eq:def_y_function1}, \eqref{eq:def_y_function2}. Note that a crucial ingredient for this is the knowledge of the functions $\Phi_1^{(1)}(\lambda)$ and $\Phi_1^{(2)}(\lambda)$ obtained from the explicit form of the $T$-system.

Having constructed the $Y$-functions $y^{(1)}_{1}(\lambda)$ and $y^{(2)}_{1}(\lambda)$ for finite $N$, we can obtain numerically all the higher $Y$-functions through the $Y$-system, and study the corresponding analytic behavior inside the physical strip \eqref{eq:physical_strip}. Analytical inspection for finite values of $N$ reveals the following behavior
\begin{itemize}
	\item  $y^{(1)}_{1}(\lambda)$ and $y^{(2)}_{1}(\lambda)$ have a double pole at $\lambda=0$ for $j$ odd, and a double zero at $\lambda=0$ for $j$ even;
	\item in addition, $y^{(1)}_1(\lambda)$ has a pair of order $N$ zeros at $\lambda =\pm i \left( \frac{1}{2} +  \frac{\beta}{N} \right)$ and a pair of order $N$ poles at $\lambda =\pm i \left( \frac{1}{2} - \frac{\beta}{N} \right)$, where we recall that $\beta$ enters the definition of the transfer matrices through the inhomogeneities \eqref{eq:inhomogeneities_1},\eqref{eq:inhomogeneities_2}.
\end{itemize} 
The above analytical structure has been verified numerically up to $N=8$. In the following we assume it holds for all the values of $N$, as can be checked a posteriori from the explicit knowledge of the $Y$ functions, determined as a solution of a set of non-linear integral equations.  There, the poles and zeroes are determined in a self-consistent manner, and we indeed check that the analytic structure inside and in the vicinity of the physical strip is stable as $N$ increases.

Next, we note that the following asymptotic behavior of the $Y$-functions can be deduced
\be 
\lim_{\lambda\to \pm \infty} y_m^{(1)}(\lambda)=\lim_{\lambda\to \pm \infty} y_m^{(2)}(\lambda)= C_m\,,
\ee 
where $C_m$ are determined from the recursive relations: 
\be 
1+C_{m+1} = \frac{C_m(1+C_m)}{1+C_{m-1}}\,,
\label{eq:recursion}
\ee 
with the additional constraint $C_0 = 0$, and $C_1=2$. The recursion relation \eqref{eq:recursion} can be solved explicitly, yielding
\be
C_{n}=\frac{(n+1)(n+2)}{2}-1\,.
\label{eq:asymptotic}
\ee

Using the properties of the $Y$-functions stated above, it is straightforward to perform the procedure sketched in the previous section (see \cite{PiPV17} for more details). In particular, we arrive at
\bea
 \ln y_1^{(1)}(\lambda) = s \ast \ln\left(  \frac{1+y_2^{(1)}}{1+(y_1^{(2)})^{-1}} \right)(\lambda) + 2 \ln \coth\left ( \frac{\pi \lambda}{2} \right)\nonumber\\
- N \ln \frac{ \cosh(\pi \lambda) - \sin\left( {\pi \beta  \over N} \right) }{ \cosh(\pi \lambda) + \sin\left( {\pi \beta  \over N} \right) }\,,
\\
 \ln y_1^{(2)}(\lambda) = s \ast \ln\left(  \frac{1+y_2^{(2)}}{1+(y_1^{(1)})^{-1}} \right)(\lambda) + 2 \ln \coth\left ( \frac{\pi \lambda}{2} \right)\,,
\eea
and, for $m>1$, 
\bea
\fl \ln y_m^{(1)}(\lambda) = s \ast \ln\left(  \frac{(1+y_{m+1}^{(1)})(1+y_{m-1}^{(1)})}{1+(y_m^{(2)})^{-1}} \right)(\lambda) + 2(-1)^{m+1}  \ln \coth\left ( \frac{\pi \lambda}{2} \right)\,,
\\
\fl \ln y_m^{(2)}(\lambda) = s \ast \ln\left(  \frac{(1+y_{m+1}^{(2)})(1+y_{m-1}^{(2)})}{1+(y_m^{(1)})^{-1}} \right)(\lambda) + 2  (-1)^{m+1} \ln \coth\left ( \frac{\pi \lambda}{2} \right)\,,
\eea
where 
\be
s(\lambda) = \frac{1}{2 \cosh(\pi \lambda)}\,.
\label{eq:s_function}
\ee
Here we introduced the convolution between two functions
\be
\left[f\ast g\right](\lambda)=\int^{+\infty}_{-\infty}{\rm d}\mu f(\lambda-\mu)g(\mu)\,.
\ee
The above integral equations can be solved numerically to high precision. In order to do this, the infinite system has to be truncated to a maximum number of equations $n_{\rm M}$. The truncation can be performed efficiently by imposing the boundary condition
\be
y_{n_{\rm M}+1}=y_{n_{\rm M}-1}\left(1+\frac{4}{n_{\rm M}}\right)\,,
\ee
which automatically enforces the asymptotic behavior \eqref{eq:asymptotic}.

Once the $Y$-functions are computed, the eigenvalues of $\tau^s_N(\lambda)$ and $\bar{\tau}^s_N(\lambda)$ can be obtained from the definition of the $Y$-functions 
\bea 
1+y_1^{(1)}(\lambda) &=& \frac{\tau^s_N\left( \lambda+\frac{i}{2}\right)\tau^s_N\left( \lambda-\frac{i}{2}\right)}{\tilde{\Phi}^{(1)}_1(\lambda) \bar{\tau}^s_N(\lambda)}\,, \\
1+y_1^{(2)}(\lambda) &=& \frac{\bar{\tau}^s_N\left( \lambda+\frac{i}{2}\right)\bar{\tau}^s_N\left( \lambda-\frac{i}{2}\right)}{\tilde{\Phi}^{(2)}_1(\lambda) \tau^s_N(\lambda)} \,,
\eea 
where 
\bea 
\tilde{\Phi}^{(1)}_1(\lambda) &=&  \frac{\lambda^2}{\lambda^2 + \frac{1}{4}}
\left[\left(\lambda + i\left(\frac{1}{2} +\frac{\beta}{N} \right)\right)
\left(\lambda- i\left(\frac{1}{2}+\frac{\beta}{N} \right)  \right)
\right]^N  \,,\\
\tilde{\Phi}^{(2)}_1(\lambda) &=& \frac{\lambda^2}{\lambda^2 + \frac{1}{4}}
\left[\left(\lambda + i\left(2 -\frac{\beta}{N} \right)\right)
\left(\lambda- i\left(2 -\frac{\beta}{N} \right)  \right)
\right]^N \,.
\eea 
In fact, in the following we will be interested in the leading eigenvalue of $\mathcal{T}$, defined in \eqref{eq:key_relation}. Hence, it is convenient to consider the normalized eigenvalues $\Lambda(\lambda)$ and  $\bar{\Lambda}(\lambda)$ 
\bea  
 \Lambda(\lambda) &=&  \left[ \left( \lambda + i \left(  -\frac{\beta}{N}+1 \right) \right) 
\left( \lambda - i \left( -\frac{\beta}{N}+1 \right) \right) 
\right]^{-N}\tau^s_N(\lambda)\,, \\
\bar{\Lambda}(\lambda) &=&  \left[ \left( \lambda + i \left( - \frac{\beta}{N}+\frac{3}{2} \right) \right) 
\left( \lambda - i \left( - \frac{\beta}{N}+\frac{3}{2} \right) \right) 
\right]^{-N} \bar{\tau}^s_N(\lambda)\,.
\eea
The function $\Lambda(\lambda)$, by construction, is such that $(1/3)\Lambda(0)$ is the leading eigenvalue of $\mathcal{T}$ (the prefactor $1/3$ takes into account the normalization of the initial delta-state). It is now straightforward to rewrite
\bea 
1+y_1^{(1)}(\lambda) &=& \frac{\lambda^2+ \frac{1}{4}}{\lambda^2}
\frac{\Lambda\left( \lambda+\frac{i}{2}\right)\Lambda\left( \lambda-\frac{i}{2}\right)}{ \bar{\Lambda}(\lambda)}\,,
\\  
1+y_1^{(2)}(\lambda) &=&
\left[
\frac{ 
	\left( \lambda^2 + \left( \frac{\beta}{N}-2\right)^2\right)
}
{
	\left(\lambda^2 + \left(\frac{\beta}{N} +2\right)^2\right) }
\right]^N
\frac{\lambda^2+ \frac{1}{4}}{\lambda^2}
\frac{\bar{\Lambda}\left( \lambda+\frac{i}{2}\right)\bar{\Lambda}\left( \lambda-\frac{i}{2}\right)}{ {\Lambda}(\lambda)}\,.
\eea 
We now make the assumption that $ {\Lambda}(\lambda)$ and $\bar{\Lambda}(\lambda)$ have no poles inside the physical strip, which we verified numerically for sizes up to $N=8$ and can once again be checked by examining the numerical solutions the non-linear integral equations. Introducing 
\be 
\psi_N(\lambda)= \ln \frac{ 
	\left( \lambda^2 + \left( \frac{\beta}{N}-2\right)^2\right)
}
{
	\left(\lambda^2 + \left(\frac{\beta}{N}+2\right)^2\right) } \,,
\ee 
we finally arrive at the non-linear integral equation
\bea 
\ln \Lambda &=& s\ast \ln(1+y_1^{(1)})  + s\ast \log \bar{\Lambda} - s \ast\ln \frac{\lambda^2+ \frac{1}{4}}{\lambda^2}  \\
\ln \bar{\Lambda} &=& s\ast \ln(1+y_1^{(2)})  + s\ast \log{\Lambda} - s \ast\ln \frac{\lambda^2+ \frac{1}{4}}{\lambda^2}   -  N s \ast \psi_N\,.
\eea 

As a final step, we modify the integral equations above in such a way that $\ln \Lambda(\lambda)$ and $\ln \bar{\Lambda}(\lambda)$ are explicitly expressed in terms of $y^{(r)}_1$ with $r=1,2$. Following \cite{AnFL83}, we define
\bea
Q^{(1)}(\lambda)&=&\ln \Lambda(\lambda)\,,\\
Q^{(2)}(\lambda)&=&\ln \overline{ \Lambda }(\lambda)\,,
\eea
so that we can rewrite the equations as
\be
s^{-1}\left[Q^{(r)}(\lambda)\right]=Q^{(\bar{r})}(\lambda)+h^{(r)}(\lambda)\,,
\label{eq:to_solve}
\ee
where $r=1,2$ while
\be
\bar{r}=
\left\{
\begin{array}{ll}
	2 & {\rm if \ } r=1\,,\\
	1 & {\rm if \ } r=2\,,
\end{array}
\right.
\ee
and
\bea
h^{(1)}(\lambda)&=& \ln \left( 1 + y _ { 1 } ^ { ( 1 ) }(\lambda) \right) - \ln \frac { \lambda ^ { 2 } + \frac { 1 } { 4 } } { \lambda ^ { 2 } } \,,\\
h^{(2)}(\lambda)&=& \ln \left( 1 + y _ { 1 } ^ { ( 2 ) } (\lambda)\right) - \ln \frac { \lambda ^ { 2 } + \frac { 1 } { 4 } } { \lambda ^ { 2 } }- N \psi_N(\lambda) \,.
\eea
Here we introduced the operator $s^{-1}$ which is understood to act in Fourier space as multiplication by $\hat{s}^{-1}(k)$ [where $s(\lambda)$ is defined in \eqref{eq:s_function}, and where $\hat{s}(k)$ denotes the Fourier transform of $s(\lambda)$]. A formal solution to Eq.~\eqref{eq:to_solve} can be written down explicitly (see \cite{MBPC17}). In particular, one has
\be
Q^{(1)}(\lambda)=\left[\mathcal{G}_{2}\ast h^{(1)}\right](\lambda)+\left[\mathcal{G}_{1}\ast h^{(2)}\right](\lambda)\,,
\ee 
where
\bea
\mathcal{G}_{1}(\lambda)&=&\frac { 1 } { \sqrt { 3 } } \frac { 1 } { ( 2 \cosh ( 2 \pi \lambda / 3 ) + 1 ) }\,,\\
\mathcal{G}_{2}(\lambda)&=&\frac { 1 } { \sqrt { 3 } } \frac { 1 } { ( 2 \cosh ( 2 \pi \lambda / 3 ) - 1 ) }\,.
\eea
Writing down everything explicitly we obtain the final result 
\bea
\ln \Lambda(\lambda)&=& +\left[\mathcal{G}_{2}\ast \ln\left(1+y^{(1)}_1\right)\right](\lambda) +\left[\mathcal{G}_{1}\ast \ln\left(1+y^{(2)}_1\right)\right](\lambda)\nonumber\\
&-&\left[(\mathcal{G}_{1}+\mathcal{G}_{2})\ast \log\frac{\lambda^2+1/4}{\lambda^2}\right](\lambda)-N\left[\mathcal{G}_{1}\ast \psi_N\right](\lambda)\,.
\eea

The results presented so far give us access to the leading eigenvalue of the boundary transfer matrix $\tau^s_N(\lambda)$ for arbitrary length $N$ of the chain. However, it is also possible to compute the limit $N\to\infty$: this can be simply done by taking the limit of the $N$-dependent driving terms, which can be easily performed. Explicitly, the relevant equations for the leading eigenvalue of $\Lambda(\lambda)$ in the limit $N\to\infty$ can be summarized as follows
\bea
\fl \ln y_m^{(1)}(\lambda) = -\frac{2\pi \beta}{\cosh(\lambda\pi)}\delta_{m,1}+ s \ast \ln\left(  \frac{(1+y_{m+1}^{(1)})(1+y_{m-1}^{(1)})}{1+(y_m^{(2)})^{-1}} \right)(\lambda) \nonumber\\
+ 2(-1)^{m+1}  \ln \coth\left ( \frac{\pi \lambda}{2} \right)\,, \label{eq:final_result_QTM_1} \\
\fl \ln y_m^{(2)}(\lambda) = s \ast \ln\left(  \frac{(1+y_{m+1}^{(2)})(1+y_{m-1}^{(2)})}{1+(y_m^{(1)})^{-1}} \right)(\lambda) + 2  (-1)^{m+1} \ln \coth\left ( \frac{\pi \lambda}{2} \right)\,,\label{eq:final_result_QTM_2} 
\eea
with the convention $y^{(r)}_0(\lambda)\equiv 0$ and
\bea
\ln \Lambda(\lambda)&=& +\left[\mathcal{G}_{2}\ast \ln\left(1+y^{(1)}_1\right)\right](\lambda) +\left[\mathcal{G}_{1}\ast \ln\left(1+y^{(2)}_1\right)\right](\lambda)\nonumber\\
&-&\left[(\mathcal{G}_{1}+\mathcal{G}_{2})\ast \log\frac{\lambda^2+1/4}{\lambda^2}\right](\lambda)-\left[\mathcal{G}_{1}\ast \psi_{\infty}\right](\lambda)\,,
\label{eq:final_result_lambda}
\eea
where
\be
\psi_{\infty}(\lambda)=\frac{-8\beta}{4+\lambda^2}\,.
\ee

In Sec.~\ref{sec:lochmidt} we will present an explicit numerical solution to the above equations. Indeed, as we will see, the value of $\Lambda(0)$ is closely related to the Loschmidt echo in the thermodynamic limit. In the next subsection, instead, we will show how the solution of the integral equations above can be obtained analytically in the limit $\beta\to 0$.

\subsection{The analytic solution to the $Y$-system at vanishing times} 
\label{sec:QTM_QAM}

From the discussion of the previous section, it follows that in order to determine the solution to the $Y$-system, one needs in general to resort to a numerical evaluation of the corresponding integral equations. This is true, for example, for boundary transfer matrices associated to a given integrable state, in the case of generic values of the parameter $\beta$ parametrizing the inhomogeneities [cf. Eq.~\eqref{eq:inhomogeneities_2}]. However, it is possible to derive an analytic solution to the $Y$-system in the limit $\beta\to 0$. This result is particularly important due to its connection to the problem of deriving the quasi-momentum distribution functions of the quasi-particles in the post-quench steady state \cite{PVCP18}. In this section, we present such an analytical solution for arbitrary integrable states.

Following \cite{PiPV17}, we start by introducing the operators
\bea
\mathcal{U}(\lambda)&=&\frac{\langle\psi_0|T_a(-\lambda)\otimes T_a(\lambda)|\psi_0\rangle}{(\lambda+i)^N(\lambda-i)^N}\,,\\
\widetilde{\mathcal{U}}(\lambda)&=&\frac{\langle\tilde{\psi}_0|T_{\bar{a}}(-\lambda)\otimes T_{\bar{a}}(\lambda)|\tilde{\psi}_0\rangle}{(\lambda+3i/2)^N(\lambda-3i/2)^N}\,,
\label{eq:mathcalT_lambda}
\eea
where $T_a(\lambda)$ and $T_{\bar{a}}(\lambda)$ are defined in \eqref{eq:shifted_monodromy} and \eqref{eq:shifted_monodromy_bar} respectively, while $|\psi_0\rangle$ is given in \eqref{eq:integrable_block}. Here we also defined
\bea 
|\tilde{\psi}_0 \rangle &=& \tilde{k}_{11} |1,1\rangle +\tilde{k}_{22} |2,2\rangle +  \tilde{k}_{33} |3,3\rangle 
+ \tilde{k}_{12}( |1,2\rangle +  |2,1\rangle )\nonumber\\ 
&+& \tilde{k}_{13}( |1,3\rangle +  |3,1\rangle ) 
+ \tilde{k}_{23}( |2,3\rangle +  |3,2\rangle )\,,
\label{eq:inverse_integrable_block}
\eea 
where $\tilde{k}_{ij}$ are the entries of the inverse (symmetric) matrix $K_s^{-1}$, namely
\be
K_{s}^{-1}=\left(
\begin{array}{ccc}
	\tilde{k}_{11} & \tilde{k}_{12} & \tilde{k}_{13}\\
	\tilde{k}_{12} & \tilde{k}_{22} & \tilde{k}_{23}\\
	\tilde{k}_{13} & \tilde{k}_{23} & \tilde{k}_{33}
\end{array}
\right)\,.
\ee
Note that one has the identification
\bea
\mathcal{U}(\lambda)&=&\frac{1}{\mathcal{N}}\frac{\tau^s_N(\lambda)}{(\lambda+i)^N(\lambda-i)^N}\,,\label{eq:def_mathcalU1}\\
\widetilde{\mathcal{U}}(\lambda)&=&\frac{1}{\widetilde{\mathcal{N}}}\frac{\bar{\tau}^s_N(\lambda)}{(\lambda+3i/2)^N(\lambda-3i/2)^N}\,,\label{eq:def_mathcalU2}
\eea
where $\mathcal{N}=\langle\psi_0|\psi_0\rangle$ and $\widetilde{\mathcal{N}}=\langle\tilde{\psi}_0|\tilde{\psi}_0\rangle$, and where $\tau^s_N(\lambda)$, $\bar{\tau}^s_N(\lambda)$ are defined in \eqref{eq:open_transfer_matrix}, \eqref{eq:bar_open_transfer_matrix}.
Now, using simple manipulations, it is not difficult to show
\bea
\fl \lim_{\beta\to 0}\frac{ {\rm tr}\left\{\left[{}_{\mathcal{N}}{\langle \psi |T_a(\lambda)\otimes T_a(-\lambda) |\psi\rangle}{_{\mathcal{N}}}\right]^{L/2}\right\}}{{\left[(\lambda+i)(\lambda-i)\right]^{NL}}}&=&\frac{\langle \Psi^+_0 | \left[ t_{\lambda,-\lambda}(0) \bar{t}_{\lambda,-\lambda}(0) \right]^N  |  \Psi^-_0\rangle }{\left[(\lambda+i)(\lambda-i)\right]^{NL}}\,,\\
\fl \lim_{\beta\to 0}\frac{ {\rm tr}\left\{\left[{}_{\mathcal{N}}{\langle \tilde{\psi} |T_{\bar{a}}(\lambda)\otimes T_{\bar{a}}(-\lambda) |\tilde{\psi}\rangle}{_{\mathcal{N}}}\right]^{L/2}\right\}}{{\left[(\lambda+3i/2)(\lambda-3i/2)\right]^{NL}}}&=&\frac{\langle \tilde{\Psi}^+_0 | \left[ t_{\lambda,-\lambda}(-3i/2) \bar{t}_{\lambda,-\lambda}(3i/2) \right]^N  |\tilde{\Psi}^-_0\rangle }{{\left[(\lambda+3i/2)(\lambda-3i/2)\right]^{NL}}}\,,
\eea
where we defined
\bea
|\Psi^{-}_0\rangle=|\psi_0\rangle_{\mathcal{N}}^{\otimes L/2}=\frac{|\psi_0\rangle^{\otimes L/2}}{\langle\psi_0|\psi_0\rangle^{L/4}}\,,\\
(|\Psi^{+}_0\rangle)^{\ast}=\left[|\psi_0\rangle_{\mathcal{N}}^{\otimes L/2}\right]^{\ast}=\left[\frac{|\psi_0\rangle^{\otimes L/2}}{\langle\psi_0|\psi_0\rangle^{L/4}}\right]^{\ast}\,,
\eea
and analogously for $|\tilde{\Psi}_0^{\pm}\rangle$ and $|\tilde{\psi}\rangle_{\mathcal{N}}$.
Here we introduced the following transfer matrices $t_{\lambda,-\lambda}(w)$ acting in the original time direction
\bea
\fl t_{\lambda,-\lambda}(w) ={\rm tr}_0 \left\{R_L(w+\lambda)R_{L-1}(w-\lambda) \ldots R_2(w+\lambda)R_1(w-\lambda)\right\}
\label{eq:transfermatrixt_dis}\\
\fl \bar{t}_{\lambda,-\lambda}(w) = {\rm tr}_0 \left\{\bar{R}_L(w+\lambda-3i/2)\bar{R}_{L-1}(w-\lambda-3i/2)\right.\nonumber\\
\left.\ldots \bar{R}_2(w+\lambda-3i/2)\bar{R}_1(w-\lambda-3i/2)\right\} \,.
\label{eq:bar_transfermatrixt_dis}
\eea   
It follows that
\bea
\frac{\langle \Psi^+_0 | \left[ t_{\lambda,-\lambda}(0) \bar{t}_{\lambda,-\lambda}(0) \right]^N  |  \Psi^-_0\rangle}{\left[(\lambda+i)(\lambda-i)\right]^{NL}}=\lim_{\beta\to 0}\tr\left\{\mathcal{U}(\lambda)\right\}^{L/2}\,,\\
\frac{\langle \Psi^+_0 | \left[ t_{\lambda,-\lambda}(-3i/2) \bar{t}_{\lambda,-\lambda}(3i/2) \right]^N  |  \Psi^-_0\rangle}{{\left[(\lambda+3i/2)(\lambda-3i/2)\right]^{NL}}}=\lim_{\beta\to 0}\tr\left\{\widetilde{\mathcal{U}}(\lambda)\right\}^{L/2}\,.
\eea

Let $\Lambda_0(\lambda)$ and $\widetilde{\Lambda}_0(\lambda)$ be the eigenvalues of $\mathcal{U}(\lambda)$, $\widetilde{\mathcal{U}}(\lambda)$ such that they are the leading one for $\lambda=0$. Then, for small values of $\lambda$ and large system sizes $L$, we have
\bea
\tr\left\{\mathcal{U}(\lambda)\right\}^{L/2}&\simeq &\Lambda_0(\lambda)^{L/2}\,,\\
\tr\left\{\widetilde{\mathcal{U}}(\lambda)\right\}^{L/2}&\simeq &\tilde{\Lambda}_0(\lambda)^{L/2}\,.
\eea
On the other hand, the following inversion relations hold for small values of $\lambda$
\bea
\lim_{L\to\infty}\frac{t_{\lambda,-\lambda}(0)\bar{t}_{\lambda,-\lambda}(0)}{\left[(\lambda+i)(\lambda-i)\right]^{L}}=\mathbf{1}\,,\label{eq:inversion_rel_1}\\
\lim_{L\to\infty}\frac{t_{\lambda,-\lambda}(-3i/2)\bar{t}_{\lambda,-\lambda}(3i/2)}{\left[(\lambda+3i/2)(\lambda-3i/2)\right]^{L}}=\mathbf{1}\,.
\label{eq:inversion_rel_2}
\eea
The rigorous meaning of \eqref{eq:inversion_rel_1} and
\eqref{eq:inversion_rel_2} and their proof is reported in \ref{sec:inversion:relations}. Putting everything together, we obtain
\be
\lim_{\beta\to 0}\Lambda_0(\lambda)\equiv \lim_{\beta\to 0} \widetilde{\Lambda}_0(\lambda)\equiv 1\,.
\ee

We now have all the ingredients to compute the analytic solution to the $y$-system in the limit $\beta\to 0$. First, it follows from the definitions \eqref{eq:def_y_function1}, \eqref{eq:def_y_function2} that
\bea
1+y_{1}^{(1)}(u)&=&\frac{\tau_N^s\left(u+\frac{i}{2}\right)\tau_N^s\left(u-\frac{i}{2}\right)}{\tilde{\Phi}^{(1)}_1(u)\bar{\tau}_N^s(u)}\,,\label{eq:y1}\\
1+y_{1}^{(2)}(u)&=&\frac{\bar{\tau}_N^s\left(u+\frac{i}{2}\right)\bar{\tau}_N^s\left(u-\frac{i}{2}\right)}{\tilde{\Phi}^{(2)}_1(u)\tau_N^s(u)}\,. \label{eq:y2}
\eea
Exploiting the relations \eqref{eq:def_mathcalU1}, \eqref{eq:def_mathcalU2}, between $\mathcal{U}(\lambda)$, $\widetilde{\mathcal{U}}(\lambda)$ and $\tau^s_N(\lambda)$, $\bar{\tau}^s_N(\lambda)$, using the results \eqref{eq:final_result_1}--\eqref{eq:final_result_4}, and collecting all the factors, we arrive at the final result, which is extremely simple and reads
\bea
\lim_{\beta\to 0}\left(1+y_{1}^{(1)}(\lambda)\right)&=&\frac{\mathcal{N}^2}{\widetilde{\mathcal{N}}({\rm det}K_s)^2}\frac{\lambda^2+1/4}{\lambda^2}\,,\label{eq:final_y1}\\
\lim_{\beta\to 0}\left(1+y_{1}^{(2)}(\lambda)\right)&=&\frac{\widetilde{\mathcal{N}}^2({\rm det}K_s)^2}{\mathcal{N}}\frac{\lambda^2+1/4}{\lambda^2}\,.\label{eq:final_y2}
\eea
Note that these expressions do not depend on $N$, and that higher $Y$-functions can be obtained using the $Y$-system relations. In the case of diagonal reflection matrices $K_s$, the above expressions can be explicitly written as
\bea
\fl \lim_{\beta\to 0} \left(1+y_{1}^{(1)}(\lambda)\right)=\frac{(|\kappa_{11}|^2+|\kappa_{11}|^2+|\kappa_{33}|^2)^2}{(1/|\kappa_{11}|^2+1/|\kappa_{22}|^2+1/|\kappa_{33}|^2)|\kappa_{11}\kappa_{22}\kappa_{33}|^2}\frac{\lambda^2+1/4}{\lambda^2}\,,\label{eq:analytical_eta1}\\
\fl \lim_{\beta\to 0} \left(1+y_{1}^{(2)}(\lambda)\right)=\frac{((1/|\kappa_{11}|^2+1/|\kappa_{22}|^2+1/|\kappa_{33}|^2)^2|\kappa_{11}\kappa_{22}\kappa_{33}|^2}{|\kappa_{11}|^2+|\kappa_{11}|^2+|\kappa_{33}|^2}\frac{\lambda^2+1/4}{\lambda^2}\,,\label{eq:analytical_eta2}
\eea

\subsection{The Loschmidt echo}
\label{sec:lochmidt}
\begin{figure}
	\centering
	\begin{tabular}{ll}
		\includegraphics[scale=0.65]{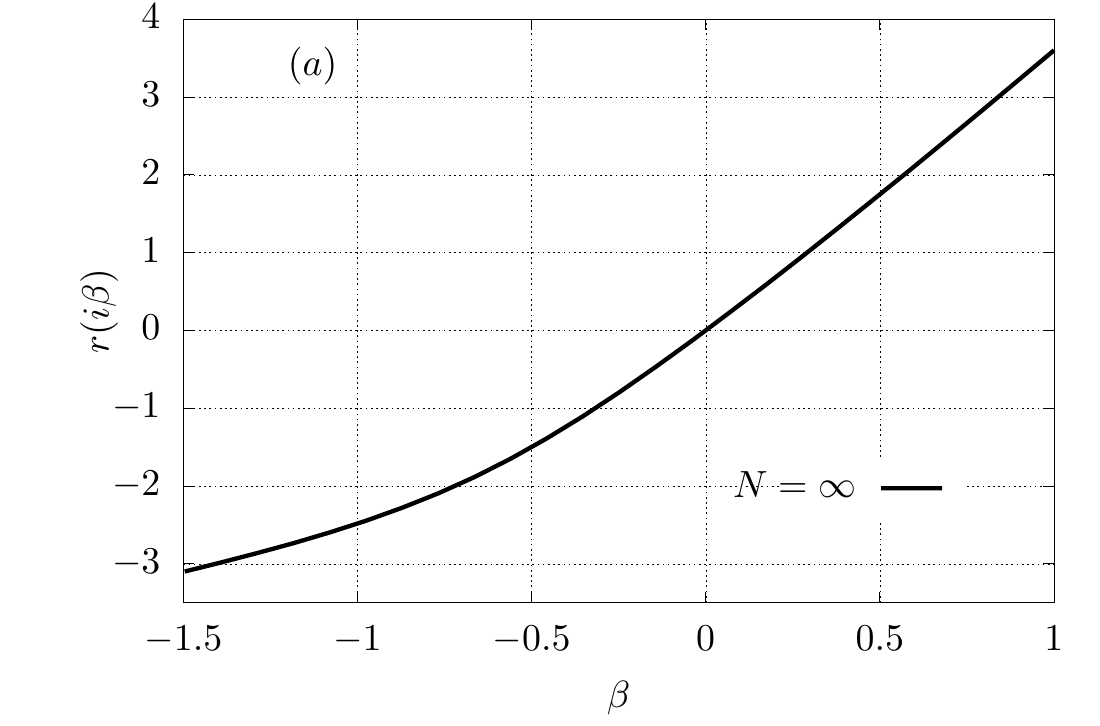} & 
		\hspace{0.25cm}	\includegraphics[scale=0.65]{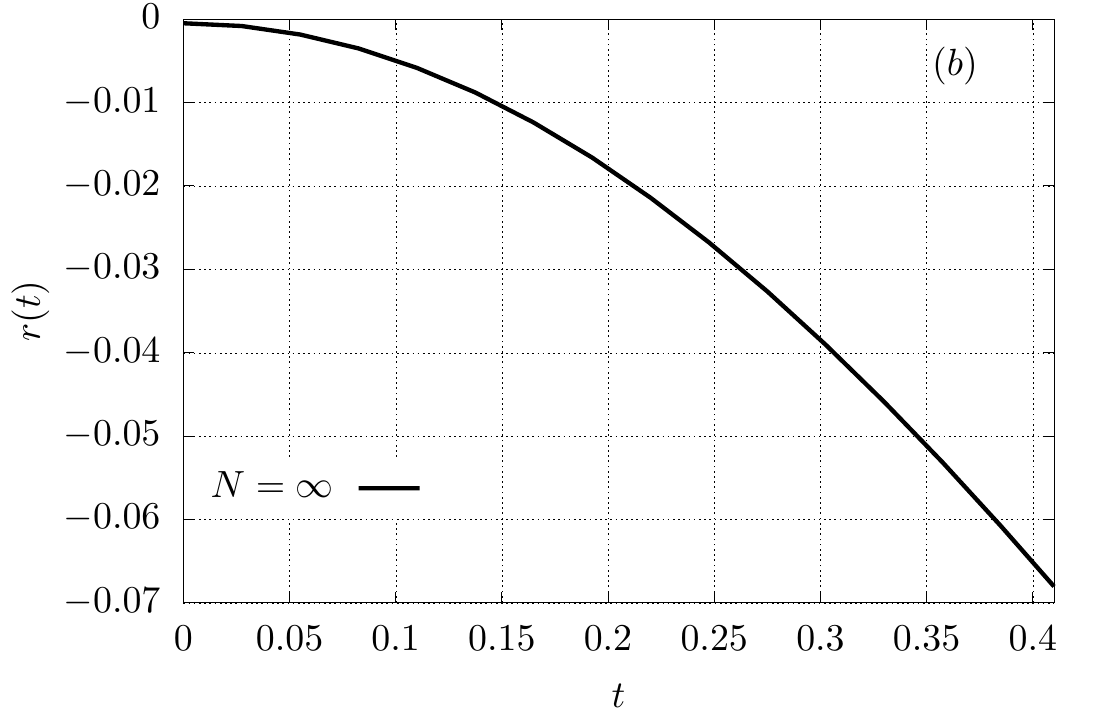}
	\end{tabular}
	\caption{Return probability for imaginary [subfigure $(a)$] and real [subfigure $(b)$] times, in the thermodynamic limit. The plots are obtained by numerical evaluation of the final result \eqref{eq:imaginary_time_return}. Finite-$N$ data, obtained by exact diagonalization calculations, are also reported for a comparison in subfigure $(b)$, showing small finite-$N$ effects for the time window considered.}
	\label{fig:loschmidt}
\end{figure}

In this section we finally present an immediate physical application of the results derived above, namely the computation of the Loschmidt echo after a quench from integrable states. While we present numerical evaluation for the case of the delta-state \eqref{eq:deltastate}, we stress that the treatment is general, and analogous calculations hold for arbitrary integrable states.

We recall that the Loschmidt echo is defined by
\be
\mathcal{L}(t)=|\langle\Psi_0|e^{iHt}|\Psi_0\rangle|^{2}\,,
\ee
and gives us information about the probability of finding the system at the initial configuration after a time $t$ after the quench.
For global quenches, the latter vanishes exponentially with the system size, so it is convenient to define the return probability
\be
r(t)=\lim_{L\to\infty}\frac{1}{L}\log \mathcal{L}(t)=\lim_{L\to\infty}\frac{2}{L}{\rm Re}\left[\log \langle\Psi_0|e^{iHt}|\Psi_0\rangle\right]\,.
\label{eq:def_return}
\ee

The Loschmidt echo is simply related to the partition function \eqref{eq:euclidean_partition_function} by $\mathcal{L}(t)=|\mathcal{Z}(it)|^2$. Furthermore, from \eqref{eq:key_relation0} one can see that the return probability \eqref{eq:def_return} can be obtained from the leading eigenvalue $\Lambda(0)$ of the transfer matrix $\mathcal{T}$ defined in \eqref{eq:key_relation}. In the case of the delta-state \eqref{eq:deltastate}, we can use directly the results of the previous section and obtain
\be
r(i\beta)=\log\Lambda(0)-\log 3\,,
\label{eq:imaginary_time_return}
\ee
where the term $\log(3)$ comes from the normalization of the delta-state. The leading eigenvalue $\Lambda(0)$ can be computed using the formula \eqref{eq:final_result_lambda}, where the $Y$-functions are obtained by solving the integral equations \eqref{eq:final_result_QTM_1}, \eqref{eq:final_result_QTM_2}. We evaluated numerically these formulas, and reported our results in Fig.\ref{fig:loschmidt}. In particular, the return probability \eqref{eq:imaginary_time_return} at imaginary times is displayed in subfigure $(a)$ of Fig.~\ref{fig:loschmidt}. We see that the behavior is the expected one: the curve is monotonic and for large values of $\beta$ we have a linear growth. The slope of the curve for large $\beta$ can in principle be computed along the lines of \cite{Pozs13}, and is expected to be expressed in terms of both the ground-state energy of the Hamiltonian \eqref{eq:hamiltonian} and the overlap between the ground state and the initial state.  

Next, following the same steps outlined in the previous section, it is also possible to compute the return probability for real times. This problem was tackled for the spin-$1/2$ XXZ chain in \cite{PiPV18}, where it was shown that several complications arise. In particular, in this case the spectrum of the boundary transfer matrix $\tau^s_N(\lambda)$ might display level crossings, which are responsible for the emergence of non-analytic points in the real-time dynamics of the return probability. In order to locate these crossings, one may examine the finite-$N$ spectrum of the QTM from exact diagonalization. In the present case, we checked that the first crossing occurs at a time $t \sim 1.5$, which guarantees in particular that the return probability remains analytic in the time window considered in Figure \ref{fig:loschmidt} (b). This approach has to be compared with that of \cite{as-14}, where the QTM is diagonalized numerically directly in the $N\to \infty$ limit. Furthermore, following the dynamics of a single eigenvalue, one finds that the analytic properties of the corresponding $Y$-functions might change in time, so that one is forced to employ an ``excited-state'' thermodynamic Bethe ansatz approach \cite{KlPe93,JuKS98, DoTa96,BaLZ97}.  We mention that while non-analytic points in the Loschmidt echo have been extensively studied in the context of dynamical quantum phase transitions \cite{hpk-13,pmgm-10,fagotti2-13,dpfz-13,ks-13,ks-17,
kks-17,cwe-14,heyl-14,vths-13,as-14,deluca-14,heyl-15,
kk-14,vd-14,ssd-15,JK-15,AbK-16,sdpd-16,zhks-16,zy-16,ps-16,Heyl17,
JaJo17,HaZa17,ZaHa17} the QTM approach provided one of the few analytical studies of the latter in the presence of interactions \cite{PiPV18}.

It follows from the above discussion that for real times the expression \eqref{eq:imaginary_time_return} (after substitution $i\beta\to t$) will be correct only up to a time $t^{\ast}$, after which the correct formula must be modified using some ``excited-state'' thermodynamic Bethe ansatz techniques \cite{KlPe93,JuKS98, DoTa96,BaLZ97}. In the $SU(3)$-invariant spin chain, this can be done in the same way explained in \cite{PiPV18} for the $XXZ$ spin-$1/2$ model. The implementation of the method goes beyond the scope of the present paper; here we simply evaluated the real-time return probability for small times, for which Eq.~\eqref{eq:imaginary_time_return} can be used. The result is plotted in subfigure $(b)$ of Fig.~\ref{fig:loschmidt}. We explicitly checked the latter against numerical calculations based on exact diagonalization for small system sizes $L$, finding perfect agreement.

\section{Conclusions}
\label{sec:conclusions}

This is the second one of two papers devoted to the study of quantum quenches in the nested $SU(3)$-invariant spin chain \eqref{eq:hamiltonian}. In this and in the previous work \cite{PVCP18}, we have generalized the QTM approach, initially developed for XXZ Heisenberg chains \cite{PiPV17,PiPV17_2}, to quantum quenches in nested systems. In particular, we have shown how it is possible to define a class of product integrable states, and how to compute exactly several quantities after the quench.

In this technical work we have focused on the derivation of some fusion relations of integrable boundary transfer matrices, which were required in order to derive the results of \cite{PVCP18}, and which are an important ingredient in the application of the QTM approach to quantum quenches. In particular, in this work we have explicitly exploited them to obtain exact formulas for the real- and imaginary-time Loschmidt echo after a quench from integrable states. We believe that the technical tools derived in this paper are interesting per se, and might find applications in other problems, possibly related to equilibrium physics in the presence of open boundary conditions.

This work, together with \cite{PVCP18}, provides a comprehensive picture for quantum quenches from integrable states, complementing the works \cite{PiPV17,PiPV17_2} where this formalism was initially developed. Furthermore, the successful application of these methods to the complicated nested models shows that integrable states are indeed ideal probes to explore quite generally the quench-dynamics in integrable systems. Finally, these works also open several directions to be investigated. Among these, one of the most fascinating questions pertains the study of quantum quenches in nested, multi-component Fermi and Bose gases, which are of great relevance for cold-atom realizations \cite{BlDZ08,GuBL13,PMCL14}.

\section*{Acknowledgments}

PC acknowledges support from ERC under Consolidator grant  number
771536 (NEMO). EV acknowledges support by the EPSRC under grant
EP/N01930X. BP was supported by the BME-Nanotechnology FIKP grant of
EMMI (BME FIKP-NAT), by the National Research Development and
Innovation Office (NKFIH) (K-2016 grant no. 119204 and KH-17 grant no. 125567), and by the
“Premium” Postdoctoral Program of the Hungarian Academy of
Sciences. Part of this work has been carried out during a visit of BP
and EV to SISSA whose hospitality is kindly acknowledged.

\appendix

\section{Inversion relations of transfer matrices} \label{sec:inversion:relations}

In this section  we will prove 
\bea
\left|\left|\frac{t_{\lambda,-\lambda}(0)\bar{t}_{\lambda,-\lambda}(0)}{\left[(\lambda+i)(\lambda-i)\right]^{L}}-\mathbf{1}\right|\right|_{\rm HS}\sim e^{-\alpha L}\,,\label{eq:inversion_to_prove_1}\\
\left|\left|\frac{\bar{t}_{\lambda,-\lambda}(3i/2)t_{\lambda,-\lambda}(-3i/2)}{\left[(\lambda+3i/2)(\lambda-3i/2)\right]^{L}}-\mathbf{1}\right|\right|_{\rm HS}\sim e^{-\beta L}\,,\label{eq:inversion_to_prove_2}
\eea
where we introduced the Hilbert-Schmidt norm of an operator as
\be
||A||_{\rm HS}=\frac{1}{3^L}{\rm tr}\left[A^{\dagger}A\right]\,,
\ee
and $\alpha, \beta>0$ are real numbers. We follow analogous calculations carried out for example in \cite{PiVe16}. We derive in particular \eqref{eq:inversion_to_prove_1}, as \eqref{eq:inversion_to_prove_2} can be established completely analogously.

We define $L(x)$, $M(x)$ via
\bea
R(x)&=&\sum_{a,b=1}^{3}L_{a,b}(x)\otimes E^{a,b}\,,\\
\bar{R}(x)&=&\sum_{a,b=1}^{3}M_{a,b}(x)\otimes E^{a,b}\,,
\eea
where we introduced the matrices $E^{a,b}$ whose matrix elements are given by
\be
(E^{a,b})_{c,d}=\delta_{a,c}\delta_{b,d}\,.
\ee
From the definitions, we have
\bea
\fl \bar{t}_{\lambda,-\lambda}(0)=\sum_{\{a\},\{b\}}{\rm tr}\left[M_{a_L,b_L}(\lambda-3i/2)M_{a_{L-1},b_{L-1}}(-\lambda-3i/2)\ldots\right]E_L^{a_L,b_L}\ldots E_1^{a_1,b_1}\,,\\
t_{\lambda,-\lambda}(0)=\sum_{\{a\},\{b\}}{\rm tr}\left[L_{a_L,b_L}(\lambda)L_{a_{L-1},b_{L-1}}(-\lambda)\ldots\right]E_L^{a_L,b_L}\ldots E_1^{a_1,b_1}\,,
\eea
so that it is easy to compute
\bea
\fl \bar{t}_{\lambda,-\lambda}(0)t_{\lambda,-\lambda}(0)=\sum_{\{a\},\{d\}}{\rm tr}\left[X_{a_L,d_L}(\lambda)X_{a_{L-1},d_{L-1}}(-\lambda)\ldots\right]E_L^{a_L,d_L}\ldots E_1^{a_1,d_1}\,,\\
\fl \left[\bar{t}_{\lambda,-\lambda}(0)t_{\lambda,-\lambda}(0)\right]^{\dagger}=\sum_{\{a\},\{d\}}{\rm tr}\left[X^\ast_{a_L,d_L}(\lambda)X^\ast_{a_{L-1},d_{L-1}}(-\lambda)\ldots\right]E_L^{d_L,a_L}\ldots E_1^{d_1,a_1}\,,
\eea
where
\bea
X_{a,d}(\lambda)=\sum_{b=1}^{3}M_{a,b}(\lambda-3i/2)\otimes L_{b,d}(\lambda)\,.
\eea
The leading eigenvalue $\Gamma_0(\lambda)$ of
$\left[\sum_{d}X_{dd}(\lambda)\right]\left[\sum_{d}X_{dd}(-\lambda)\right]$ can be easily found to be
\be
\Gamma_0(\lambda)=9\left(\lambda+i\right)^2\left(\lambda-i\right)^2\,.
\ee
Furthermore, it can be argued that $\Gamma_0(\lambda)$ has a finite gap with respect to the next to leading eigenvalues in a neighborhood of $\lambda=0$. Finally, we can compute
\bea
\left[\bar{t}_{\lambda,-\lambda}(0)t_{\lambda,-\lambda}(0)\right]^{\dagger}\bar{t}_{\lambda,-\lambda}(0)t_{\lambda,-\lambda}(0)\nonumber\\
=\sum_{\{d\},\{d'\}}{\rm tr}\left[Z_{d'_L,d_L}(\lambda)Z_{d'_L,d_L}(-\lambda)\ldots \right]E_{L}^{d'_L,d_L}\ldots E_{1}^{d'_1,d_1}\,,
\eea
and so
\bea
{\rm tr}\left\{\left[\bar{t}_{\lambda,-\lambda}(0)t_{\lambda,-\lambda}(0)\right]^{\dagger}\bar{t}_{\lambda,-\lambda}(0)t_{\lambda,-\lambda}(0)\right\}\nonumber\\
={\rm tr}\left\{\left[\left(\sum_{d=1}^3Z_{dd}(\lambda)\right)\left(\sum_{d=1}^3Z_{dd}(-\lambda)\right)\right]^{L/2}\right\}\,,
\eea
where
\be
\fl  Z_{d',d}(\lambda)=\sum_{a,b,b'=1}^3M^{\ast}_{a,b}(\lambda-3i/2)\otimes L^{\ast}_{b,d}(\lambda)M_{a,b}(\lambda-3i/2)\otimes L_{b,d}(\lambda)\,.
\ee
Similar to the previous case, we find that the leading eigenvalue $\Omega_0(\lambda)$ of 
\be
\left[\sum_{d}Z_{dd}(\lambda)\right]\left[\sum_{d}Z_{dd}(-\lambda)\right]
\ee 
has a finite gap with respect to the next to leading ones, in a neighborhood of $\lambda=0$ and it reads
\be
\Omega_0(\lambda)=9\left(\lambda+i\right)^4\left(\lambda-i\right)^4\,.
\ee
Putting everything together, it is straightforward to compute
\bea
\left|\left|\frac{\bar{t}_{\lambda,-\lambda}(0)t_{\lambda,-\lambda}(0)}{(\lambda+i)^{L}(\lambda-i)^{L}}-\mathbf{1}\right|\right|_{\rm HS}=\frac{1}{[9(\lambda+i)^{4}(\lambda-i)^{4}]^{L/2}}\Omega_0(\lambda)\nonumber\\
-\frac{1}{[9(\lambda+i)^{2}(\lambda-i)^{2}]^{L/2}}\Gamma_0(\lambda)+O(e^{-\alpha L})=O(e^{-\alpha L})\,,
\eea
with $\alpha>0$ in a neighborhood of $\lambda=0$. This proves \eqref{eq:inversion_to_prove_1}.

The other relation \eqref{eq:inversion_to_prove_2} can be proven in a
similar way, by making use of the local inversion relation \eqref{inversion2}.

\end{document}